\documentclass[12pt]{article}
\makeatletter
\def\numberbysection{\@addtoreset{equation}{section}
 	\def\theequation{\thesection.\arabic{equation}}}
\makeatother

\numberbysection 

\newtheorem{lemma}{Lemma}

\newcommand{\be}{\begin{eqnarray}}
\newcommand{\ee}{\end{eqnarray}}
\newcommand{\non}{\nonumber}
\newcommand{\n}{\ensuremath{\mathcal{N}}}
\newcommand{\tr}{\mathop{\rm tr}\nolimits}

\begin{document}

\begin{titlepage}
\strut\hfill UMTG--203
\vspace{.5in}
\begin{center}

\LARGE Bulk and Boundary $S$ Matrices for the $SU(\n)$ Chain \\[1.0in]
\large Anastasia Doikou and Rafael I. Nepomechie\\[0.8in]
\large Physics Department, P.O. Box 248046, University of Miami\\[0.2in]  
\large Coral Gables, FL 33124 USA\\

\end{center}

\vspace{.5in}

\begin{abstract}
We consider both closed and open integrable antiferromagnetic chains 
constructed with the $SU(\n)$-invariant $R$ matrix.  For the closed 
chain, we extend the analyses of Sutherland and Kulish -- Reshetikhin 
by considering also complex ``string'' solutions of the Bethe Ansatz 
equations.  Such solutions are essential to describe general 
multiparticle excited states.  We also explicitly determine the 
$SU(\n)$ quantum numbers of the states.  In particular, the model has 
particle-like excitations in the fundamental representations $[k]$ of 
$SU(\n)$, with $k = 1 \,, \ldots \,, \n-1$.  We directly compute the 
complete two-particle $S$ matrices for the cases $[1] \otimes [1]$ and 
$[1] \otimes [\n-1]$.  For the open chain with diagonal boundary 
fields, we show that the transfer matrix has the symmetry $SU(l) 
\times SU(\n-l) \times U(1)$, as well as a new ``duality'' symmetry 
which maps $l \leftrightarrow \n - l$.  With the help of these 
symmetries, we compute by means of the Bethe Ansatz for particles of 
types $[1]$ and $[\n-1]$ the corresponding boundary $S$ matrices.
\end{abstract}

\end{titlepage}

\section{Introduction}

Integrable quantum spin chains are exactly solvable quantum mechanical 
models of $N$ quantum spins, of which the Heisenberg model solved by 
Bethe \cite{bethe}-\cite {faddeev/takhtajan} is the prototype.  In 
the antiferromagnetic regime, such spin chains can be regarded as 
lattice versions of corresponding integrable relativistic quantum 
field theories.  For integrable spin chains, quantities of physical 
interest (spectrum, $S$ matrix, etc.)  can be calculated exactly by 
direct means, starting from the microscopic Hamiltonian; while for the 
corresponding field theories, such exact information has been 
primarily obtained by indirect means, such as the ``bootstrap'' 
approach \cite{zamolodchikov}, \cite{ORW} and semiclassical 
approximations.

Quantum spin chains have one spatial dimension and come in two 
topologies: ``closed'' (periodic boundary conditions) and ``open''.  
The latter exhibit a rich variety of boundary phenomena, which -- for 
integrable chains -- can be investigated exactly.

There exists a systematic approach for constructing integrable quantum 
spin chains, called the Quantum Inverse Scattering Method.  (For a 
recent review, see \cite{QISM}.)  The basic building blocks for 
constructing closed chains are $R$ matrices, which are solutions of 
the Yang-Baxter equation
\be
R_{12}(\lambda)\ R_{13}(\lambda + \lambda')\ R_{23}(\lambda')
= R_{23}(\lambda')\ R_{13}(\lambda + \lambda')\ R_{12}(\lambda) \,.
\label{YB}
\ee 
For constructing open chains, one needs also $K$ matrices, which are
solutions of the boundary Yang-Baxter equation \cite{cherednik}-\cite{GZ}
\be
R_{12}(\lambda_{1}-\lambda_{2}) K_{1}(\lambda_{1}) R_{21}(\lambda_{1}+\lambda_{2})
K_{2}(\lambda_{2}) =
K_{2}(\lambda_{2}) R_{12}(\lambda_{1}+\lambda_{2}) K_{1}(\lambda_{1}) 
R_{21}(\lambda_{1}-\lambda_{2}) \,.
\label{boundaryYB}
\ee 

In this paper, we focus on integrable quantum spin chains constructed 
with the $R$ matrix \cite{yang} 
\be
R(\lambda) = {1\over \lambda + i}\left( \lambda + i {\cal P} \right) 
\label{Rmatrix}
\,,
\ee
where ${\cal P}$ is the permutation matrix
\be
{\cal P} x \otimes y = y \otimes x
\ee
for all vectors $x$ and $y$ in an $\n$-dimensional complex vector space 
$C_{\n}$.  This $R$ matrix is $SU(\n)$ invariant; i.e.,
\be
\left[ U \otimes U \,, R (\lambda) \right] = 0 
\label{invariance}
\ee
for all group elements $U$ in the defining representation of $SU(\n)$.

This paper has two main sections.  In Section 2 we consider the 
{\it closed} integrable antiferromagnetic chain with $N$ ``spins'' 
(vectors in $C_{\n}$), which has the Hamiltonian
\be 
{\cal H}_{closed} = \sum_{n=1}^{N-1} {\cal H}_{n n+1} + {\cal H}_{N 1} \,,
\label{closedHamiltonian}
\ee
where the two-site Hamiltonian ${\cal H}_{j k}$ is given by
\be
{\cal H}_{j k} = {i \over 2} {d\over d \lambda} {\cal P}_{j k} R_{j k}(\lambda) 
\Big\vert_{\lambda=0} ={1 \over 2} \left( {\cal P} - 1 \right)_{j k}  \,.
\label{twosite}
\ee
The model is $SU(\n)$ invariant, and the space of states is 
$C_{\n}^{\otimes N}$.  This a generalization of the antiferromagnetic 
Heisenberg model, which corresponds to the case $\n =2$.

This model was first investigated by Sutherland \cite{sutherland} and 
by Kulish and Reshetikhin \cite{kulish/reshetikhin}.  These authors 
determined the energy eigenstates and eigenvalues in terms of $\n-1$ 
types of roots of a system of Bethe Ansatz equations (BAE).  
Moreover, they found that the ground state corresponds to having $\n-1$ 
``filled Fermi seas''; and they studied particle-like excited states 
\footnote{Such excitations have been called ``kinks'' or ``spinons''. 
We refer to them here simply as ``particles''.}
corresponding to ``holes'' in these seas.  These analyses considered 
only real roots of the BAE.

We extend these analyses by considering also complex ``string'' 
solutions of the BAE.  Such solutions are essential to describe 
general multiparticle excited states.  Moreover, we explicitly 
determine the $SU(\n)$ quantum numbers of the states.  In particular, 
we show that the Bethe Ansatz state consisting of one hole in the 
$k^{th}$ sea ($k = 1 \,, \ldots \,, \n-1$) and no complex strings is the 
highest weight of the fundamental representation $[k]$, corresponding 
to a Young tableau with a single column of $k$ boxes, as shown in 
Figure \ref{fig1}.
\setlength{\unitlength}{12pt}
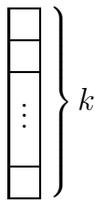
\begin{figure}[htb]
\centering
\be  
\left.
\begin{picture}(1,3.4)(0,2.8)
\put(0,0){\line(0,1){6}}
\put(1,0){\line(0,1){6}}
\put(0,0){\line(1,0){1}}
\put(0,1){\line(1,0){1}}
\put(0,4){\line(1,0){1}}
\put(0,5){\line(1,0){1}}
\put(0,6){\line(1,0){1}}
\put(.5,3.5){\makebox(0,0)[t]{$\vdots$}}
\end{picture}
\; \right\} k \non 
\ee 
\caption[xxx]{\parbox[t]{.7\textwidth}{%
 	   Young tableau with a	single column of $k$ boxes,	
 	   corresponding to	the	fundamental	representation $[k]$ of	$SU(\n)$}
 	   }
 	   \label{fig1}
\end{figure}
That is, the model has particle-like excitations in the fundamental 
representations $[k]$ of $SU(\n)$, with $k = 1 \,, \ldots \,, \n-1$.  
Finally, we directly compute the complete two-particle $S$ matrices for 
the cases $[1] \otimes [1]$ and $[1] \otimes [\n-1]$.  In the latter 
case, the singlet state is described by the Bethe Ansatz state 
consisting of two holes (one each in seas 1 and $\n-1$) as well as one 
string of length 2 in each of the $\n-1$ seas.  Our results for the $S$ 
matrices coincide with those found by the bootstrap approach 
\cite{ORW}, with no additional CDD factors.

In Section 3, we consider the {\it open} integrable chain 
constructed with the $SU(\n)$-invariant $R$ matrix (\ref{Rmatrix}), 
together with the $\n \times \n$ diagonal $K$ matrices given by 
\cite{devega/gonzalezruiz1}
\be
K^{\mp}_{(l)}(\lambda \,, \xi_{\mp}) = diag \Bigl(  
\underbrace{a^{\mp}\,, \ldots \,, a^{\mp}}_{l} \,, 
\underbrace{b^{\mp}\,, \ldots \,, b^{\mp}}_{\n-l} \Bigr) \,,
\label{Kmatrices}
\ee
where
\be
a^{-} &=& i\xi_{-} - \lambda \,, \qquad b^{-} = i\xi_{-} + \lambda \,, 
\non \\  
a^{+} &=& i\xi_{+} + \lambda \,, \qquad b^{+} = i(\xi_{+} - \n) - \lambda \,,
\label{elements}
\ee 
for any $l \in \{ 1 \,, \ldots \,, \n-1 \}$.
The Hamiltonian is given by 
\be
{\cal H}_{open} = \sum_{n=1}^{N-1} {\cal H}_{n n+1} 
+ {1\over 4\xi_{-}} {d\over d \lambda} 
K^{-}_{(l)}{}_{1}(\lambda \,, \xi_{-}) \Big\vert_{\lambda=0}
+ {\tr_{0} K^{+}_{(l)}{}_{0}(0 \,, \xi_{+}){\cal H}_{N 0}\over 
\tr K^{+}_{(l)}(0 \,, \xi_{+})} \,. 
\label{openHamiltonian}
\ee

The parameters $\xi_{\mp}$, which may be regarded as certain boundary 
magnetic fields, break the $SU(\n)$ symmetry down to $SU(l) \times 
SU(\n-l) \times U(1)$.  Moreover, we find a new ``duality'' 
symmetry which maps $l \leftrightarrow \n - l$.
With the help of these residual symmetries of the model, we compute for 
particles of types $[1]$ and $[\n-1]$ the corresponding boundary $S$ 
matrices, which describe scattering from the ends of the chain.  This 
is the first direct calculation of boundary $S$ matrices for a model 
whose symmetry algebra has rank greater than one.  For the case 
$\n=2$, we recover the results of Refs.  \cite{GMN} and \cite{DMN}.

\section{The closed chain}

In this	section, we	consider the closed	integrable chain
constructed	with the $SU(\n)$-invariant $R$ matrix (\ref{Rmatrix}).
The	transfer matrix	$t(\lambda)$ is given by
\be
t(\lambda)	= \tr_{0} T_{0}(\lambda) \,,
\ee
where $T_{0}(\lambda)$ is the monodromy	matrix
\be
T_{0}(\lambda) = R_{0N}(\lambda) \cdots  R_{01}(\lambda) \,. 
\label{monodromy}
\ee
(As is customary, we suppress the quantum-space subscripts $1 \,, 
\ldots \,, N$ of $T_{0}(\lambda)$.) The Yang-Baxter equation 
(\ref{YB}) guarantees the commutativity of the transfer matrix
\be
\left[ t(\lambda)\,, t(\lambda') \right] = 0 \,.
\ee 
The Hamiltonian (\ref{closedHamiltonian}) is proportional to the 
logarithmic derivative of the transfer matrix at $\lambda=0$ 
\be
{\cal H}_{closed} ={i \over 2}  {d\over d \lambda} \log t(\lambda)
\Big\vert_{\lambda=0} \,.
\ee 
The ``momentum'' operator $P$ is defined by 
\be
P = -i \log t(0)  \,,
\ee
since $t(0)$ is the one-site shift operator.

\subsection{$SU(\n)$ generators}

In the defining representation of $SU(\n)$, we identify the 
raising and lowering operators
\be 
j^{+ (k)} = e_{k, k+1} \,, \qquad
j^{- (k)} = e_{k+1, k} \,, \qquad k = 1 \,, \ldots \,, \n-1 \,,
\label{raising/lowering}
\ee
and the Cartan generators
\be 
s^{(k)} = e_{k, k} - e_{k+1, k+1} \,, \qquad k = 1 \,, \ldots \,, \n-1 \,,
\label{cartan}
\ee 
where $e_{k,l}$ are elementary $\n \times \n$ matrices with matrix elements 
$\left( e_{k,l} \right)_{a b} = \delta_{k,a} \delta_{l,b}$. These 
generators obey the commutation relations
\be
\left[ j^{+(k)} \,, j^{-(l)} \right] = \delta_{k,l} s^{(k)} \,, 
\qquad 
\left[ s^{(k)} \,, j^{+(l)} \right] = \left( 2 \delta_{k,l} 
-  \delta_{k,l+1} - \delta_{k+1,l} \right) j^{+(l)} \,. 
\label{algebra}
\ee 

We denote by $j^{\pm (k)}_{n}$, $s^{(k)}_{n}$ the generators at site $n$,
e.g.,
\be
s^{(k)}_{n} = 1 \otimes \ldots \otimes 1 \otimes 
\stackrel{\stackrel{n^{th}}{\downarrow}}{s^{(k)}} \otimes 1 
\otimes \ldots \otimes 1 \,, \qquad n = 1 \,, \ldots \,, N \,,
\ee
and we denote by $J^{\pm (k)}$, $S^{(k)}$ the corresponding ``total'' 
generators acting on the full space of states
\be
J^{\pm (k)} = \sum_{n=1}^{N} j^{\pm (k)}_{n}\,,  \qquad 
S^{(k)} = \sum_{n=1}^{N} s^{(k)}_{n} \,, \qquad k = 1 \,, \ldots \,, \n-1 \,.
\label{generators}
\ee 
The $SU(\n)$ invariance (\ref{invariance}) of the $R$ matrix implies 
that
\be
\left[ t(\lambda)  \,, J^{\pm (k)} \right] = 
\left[ t(\lambda)  \,, S^{(k)} \right] = 0 \,, 
\qquad k = 1 \,, \ldots \,, \n-1 \,.
\ee 

For future reference, we now relate these generators to the standard 
Cartan-Weyl basis.  We set
\be
E_{\alpha^{i}} &=& {\sqrt{2}\over 2} J^{+(i)} \,, \qquad 
E_{-\alpha^{i}} = {\sqrt{2}\over 2} J^{-(i)} \,, \qquad \,, \non \\
H_{i} &=& \sum_{j=1}^{\n-1} \mu_{i}^{j} S^{(j)} \,, \qquad 
i = 1 \,, \ldots \,, \n-1 
\label{cartan/weyl/basis}
\ee 
(with the coefficients $\mu_{i}^{j}$ still to be determined), and we 
demand the commutation relations
\be
\left[ H_{i} \,, E_{\alpha} \right] = \alpha_{i} E_{\alpha} \,, \qquad
\left[ E_{\alpha} \,, E_{-\alpha} \right] = \sum_{i=1}^{\n-1} \alpha_{i} 
H_{i} \,.
\label{cartan/weyl/algebra}
\ee
The vectors $\alpha^{i} = (\alpha^{i}_{1} \,, \ldots \,, 
\alpha^{i}_{\n-1})$ are the simple roots normalized to unity 
$\alpha^{i} \cdot \alpha^{i} = 1$. One finds
\be
\alpha^{i} = \Bigl(0 \,,  \ldots \,, 0 \,, -\sqrt{i-1\over 2i} \,,
\stackrel{\stackrel{i^{th}}{\downarrow}}
{\sqrt{i+1\over 2i}} \,, 0 \,,  \ldots \,, 0 \Bigr) \,.
\ee
From the second relation in Eq. (\ref{cartan/weyl/algebra}), we obtain
\be
\alpha^{j} \cdot \mu^{k} = {1\over 2} \delta_{j,k} \,.
\label{important}
\ee
This implies the important result that $\mu^{k} = (\mu^{k}_{1} \,, \ldots \,, 
\mu^{k}_{\n-1}) \,, \quad  k = 1 \,, \ldots \,, \n-1$ are the fundamental 
weights of $SU(\n)$ (see, e.g., \cite{georgi}). 

\subsection{Bethe Ansatz and string hypothesis}

We now review the exact Bethe Ansatz solution of the closed 
$SU(\n)$-invariant chain, and we use the string hypothesis to recast 
the BAE into a form which is particularly suitable for studying 
multiparticle excitations.

Since the operators ${\cal H}_{closed} \,, P \,, S^{(k)}$ mutually commute, 
there exist simultaneous eigenstates $| E \,, P \,, S^{(k)}\rangle$.
The so-called Bethe Ansatz states are the subset of these states
which are highest weights of $SU(\n)$, i.e.,
\be
J^{+ (k)} |\quad  \rangle = 0 \,, \qquad k = 1 \,, \ldots \,, \n-1 \,.
\ee 
(See, e.g., Refs. \cite{faddeev/takhtajan}, \cite{mn/mpla}, 
\cite{devega/gonzalezruiz2}.)
These states have been determined by both the coordinate \cite {sutherland} 
and algebraic \cite{kulish/reshetikhin}, \cite{devega} Bethe Ansatz 
methods. \footnote{The 
remaining states are obtained by acting on the Bethe Ansatz states
with the lowering operators $J^{- (k)}$.} In the latter approach, the Bethe Ansatz 
states are constructed using certain creation operators (elements of 
the monodromy matrix) depending on solutions $\{ \lambda_{\alpha}^{(k)} \}$ 
of the BAE 
\be
1 &=& - \prod_{\beta=1}^{M^{(k-1)}} 
e_{-1}(\lambda_{\alpha}^{(k)} - \lambda_{\beta}^{(k-1)})
\prod_{\beta=1}^{M^{(k)}} 
e_{2}(\lambda_{\alpha}^{(k)} - \lambda_{\beta}^{(k)})
\prod_{\beta=1}^{M^{(k+1)}} 
e_{-1}(\lambda_{\alpha}^{(k)} - \lambda_{\beta}^{(k+1)}) \,, \non \\
& &  \qquad \qquad \qquad 
\alpha = 1 \,, \ldots \,, M^{(k)} \,, \qquad  k = 1\,, \ldots \,, \n-1 
\,,
\label{closedBAE}
\ee
where
\be
e_n(\lambda) = {\lambda + {in\over 2}\over \lambda - {in\over 2}} \,,
\ee 
and $M^{(0)} = N \,, \quad M^{(\n)} = 0 \,, \quad 
\lambda_{\alpha}^{(0)} = \lambda_{\alpha}^{(\n)} = 0 \,.$
The corresponding eigenvalues are given by
\be
E &=& - {1\over 2} \sum_{\alpha=1}^{M^{(1)}}
{1\over {\lambda_{\alpha}^{(1)}}^{2} + {1\over 4}} \,, 
\non \\ 
P &=& -i \sum_{\alpha=1}^{M^{(1)}} \log e_{1}(\lambda_{\alpha}^{(1)}) \,,
\non  \\
S^{(k)} &=& M^{(k-1)} + M^{(k+1)} - 2 M^{(k)} \,.
\label{eigenvalues1}
\ee 

We adopt the ``string hypothesis'', which states that in the 
thermodynamic ($N \rightarrow \infty$) limit, all the solutions 
$\{ \lambda_{1}^{(k)} \,, \ldots \,, \lambda_{M^{(k)}}^{(k)} \}$ are 
collections of $M^{(n,k)}$ strings of length $n$ of the form (for 
$M^{(n,k)} > 0$)
\be
\lambda_\alpha^{(n,k,j)} = 
\lambda_\alpha^{(n,k)} + i \left({n+1\over 2} - j \right)
\,, \label{string} 
\ee
where $j = 1 \,, \ldots \,, n$; $\alpha = 1\,, \ldots \,, M^{(n,k)}$; 
$k = 1 \,, \ldots \,, \n-1$; $n = 1, \,, \ldots \,, \infty$; and the 
``centers'' $\lambda_\alpha^{(n,k)} $ are real. The total number of 
$\lambda$'s of type $k$ is given by
\be
M^{(k)} = \sum_{n=1}^\infty n M^{(n,k)} \,, \qquad 
k = 1\,, \ldots \,, \n-1 \,. 
\ee

Implementing this hypothesis in the BAE and then forming the product 
$\prod_{j=1}^n$ over the imaginary parts of the strings (see, e.g., 
\cite{faddeev/takhtajan}), we obtain a set of equations for the 
centers $\lambda_\alpha^{(n,k)}$ given (up to an overall sign) by
\be
&1& = \left\{ \prod_{m=1}^\infty \prod_{\beta=1}^{M^{(m,k-1)}} 
F_{nm}(\lambda_{\alpha}^{(n,k)} - \lambda_{\beta}^{(m,k-1)}) \right\} 
\left\{ \prod_{m=1}^\infty \prod_{\beta=1}^{M^{(m,k)}} 
E_{nm}(\lambda_{\alpha}^{(n,k)} - \lambda_{\beta}^{(m,k)}) \right\} \non \\
 &\times& \left\{ \prod_{m=1}^\infty \prod_{\beta=1}^{M^{(m,k+1)}} 
F_{nm}(\lambda_{\alpha}^{(n,k)} - \lambda_{\beta}^{(m,k+1)}) \right\} \,, 
\alpha = 1 \,, \ldots \,, M^{(n,k)} \,,   
k = 1\,, \ldots \,, \n-1 \,,
\label{BAcenters}
\ee 
where
\be
E_{nm}(\lambda) &=& e_{|n-m|}(\lambda)\ e_{|n-m|+2}^2(\lambda)\ \cdots
\ e_{n+m-2}^2(\lambda)\ e_{n+m}(\lambda) \,, \non \\
F_{nm}(\lambda) &=& e_{-(|n-m|+1)}(\lambda)\ e_{-(|n-m|+3)}(\lambda)\ \cdots
\ e_{-(n+m-3)}(\lambda)\ e_{-(n+m-1)}(\lambda) \,,
\ee
and $M^{(n,0)} = N \delta_{n,1}\,, \quad M^{(n,\n)} = 0 \,, \quad 
\lambda_{\alpha}^{(n,0)} = \lambda_{\alpha}^{(n,\n)} = 0 \,.$

Since Eqs. (\ref{BAcenters}) involve only products of phases,
it is useful to take the logarithm. We obtain the following important
equations for $\lambda_\alpha^{(n,k)}$:
\be
h^{(n,k)} ( \lambda_\alpha^{(n,k)} ) = J_\alpha^{(n,k)}  \,, 
\qquad \alpha &=& 1\,, \ldots \,, M^{(n,k)} \,, \non \\ 
k &=& 1 \,, \ldots \,, \n-1\,, \quad n = 1, \,, \ldots \,, \infty \,. 
\label{BAlog} 
\ee 
The so-called counting function $h^{(n,k)}(\lambda)$ is defined by
\be
h^{(n,k)}(\lambda) &=& {1\over 2\pi} \Bigl\{ 
- \sum_{m=1}^\infty \sum_{\beta=1}^{M^{(m,k-1)}} 
\Phi_{n m}(\lambda - \lambda_{\beta}^{(m,k-1)})
- \sum_{m=1}^\infty \sum_{\beta=1}^{M^{(m,k)}} 
\Xi_{n m}(\lambda - \lambda_{\beta}^{(m,k)}) \non \\ 
&-& \sum_{m=1}^\infty \sum_{\beta=1}^{M^{(m,k+1)}} 
\Phi_{n m}(\lambda - \lambda_{\beta}^{(m,k+1)}) \Bigr\} \,, 
\label{hn}  
\ee 
where
\be
\Xi_{n m} (\lambda) &=& (1 - \delta_{n,m}) q_{|n-m|}(\lambda) 
+ 2q_{|n-m|+2}(\lambda)
+ \cdots + 2q_{n+m-2}(\lambda) + q_{n+m}(\lambda) \,, \non \\ 
\Phi_{n m} (\lambda) &=& - \left[ q_{|n-m|+1}(\lambda) 
+ q_{|n-m|+3}(\lambda)
+ \cdots + q_{n+m-3}(\lambda) + q_{n+m-1}(\lambda) \right] \,,
\ee
and $q_n (\lambda)$ is the odd monotonic-increasing function defined 
(for $n > 0$) by
\be
q_n (\lambda) = \pi + i\log e_n (\lambda) 
\,, \qquad -\pi < q_n (\lambda) \le \pi \,.
\ee 
Moreover,
$\{ J_\alpha^{(n,k)} \}$ are integers or half-odd integers which satisfy
\be 
-J_{\max}^{(n,k)} \le J_\alpha^{(n,k)} \le J_{\max}^{(n,k)}  \,,  
\label{range}
\ee 
where $J_{\max}^{(n,k)}$ is given by
\be
J_{\max}^{(n,k)} =
{1\over 2}\left\{ M^{(n,k)} - 1 
+ \sum_{m=1}^\infty \min (m,n)\ 
\left[ M^{(m,k-1)} + M^{(m,k+1)} - 2 M^{(m,k)} \right] \right\} \,. 
\label{jmax} 
\ee 
In deriving the last equation, we assume the prescription \cite{faddeev/takhtajan}
that $J_\alpha^{(n,k)} \rightarrow J_{\max}^{(n,k)} + n$ for 
$\lambda_\alpha^{(n,k)} \rightarrow \infty$.
We further assume that $\{ J_\alpha^{(n,k)} \}$ can be regarded 
as ``quantum numbers'' of the model: for every set $\{ J_\alpha^{(n,k)} \}$ in 
the range (\ref{range}) (no two of which are identical),
there is a unique solution $\{ \lambda_\alpha^{(n,k)} \}$ (no two of which are 
identical) of (\ref{BAlog}).

Using the string hypothesis, the expressions (\ref{eigenvalues1}) for
the eigenvalues become
\be
E &=& - \pi \sum_{n=1}^\infty \sum_{\alpha=1}^{M^{(n,1)}} 
a_n (\lambda_\alpha^{(n,1)}) \,, \non \\
P &=& - \sum_{n=1}^\infty \sum_{\alpha=1}^{M^{(n,1)}} 
\left[ q_n (\lambda_\alpha^{(n,1)}) - \pi \right] \,, \non \\
S^{(k)} &=& \sum_{n=1}^\infty n \left[ 
M^{(n,k-1)} + M^{(n,k+1)} - 2 M^{(n,k)} \right] \,.  
\label{eigenvalues2} 
\ee
As already noted, $M^{(n,0)} = N \delta_{n,1}\,, \quad M^{(n,\n)} = 
0$.  

By invoking the string hypothesis, we have transformed the problem
of finding complex solutions of the BAE (\ref{closedBAE}) to the simpler
problem of finding real solutions of the equations (\ref{BAlog}).
We now proceed to discuss the ground state and excitations.

\subsection{Ground state and excitations}

One can argue\footnote{One argument is based on the observation that a 
system at temperature $T$ goes to its ground state as $T \rightarrow 
0$.  The thermodynamics for the case $\n=3$ was formulated, following 
\cite{yang/yang} -- \cite{gaudin2}, in Ref.  \cite{mntt}.} that the 
ground state is described by only real roots (i.e., strings of 
length 1) and no holes.  That is, $M^{(n,j)} = 0$ for $n>1$ and 
$M^{(1,j)} = 2 J_{\max}^{(1,j)} + 1$. Hence, $M^{(1,j)} =  N (\n - j)/\n$.
Evidently, the ground state lies in the sector where $N/\n$ is an 
integer. Moreover, this state has all $S^{(j)}=0$, and therefore is a 
singlet of $SU(\n)$, as expected for an antiferromagnet. Since there 
are no holes, this state corresponds to a set of $\n - 1$ filled Fermi seas. 

Excited states are described by root distributions with holes and 
(optionally) complex strings (i.e., strings of length greater than 1).  
We let $\nu^{(j)}$ denote the number of holes in the $j^{th}$ sea, 
\be
\nu^{(j)} = \left( 2 J_{\max}^{(1,j)} + 1 \right) - M^{(1,j)} \,.
\label{numberholes}
\ee

\vspace{.2in}
\noindent
{\bf Case a: no complex strings}
\vspace{.2in}

\noindent 
We first consider the simpler case of excited states with holes but no 
complex strings.  We refer to this as case {\bf a}.  For this case, we 
obtain from Eqs.  (\ref{jmax}) and (\ref{eigenvalues2}) the remarkably 
simple relation
\be
S^{(j)} = \nu^{(j)}  \,.
\label{nice}
\ee
It follows from Eq. (\ref{cartan/weyl/basis}) that the Cartan 
generators $H_{i}$ have the eigenvalues
\be
H_{i} = \sum_{j=1}^{\n-1} \mu_{i}^{j} \nu^{(j)} \,.
\ee
We conclude that {\it the Bethe Ansatz state with
$\nu^{(j)}$ holes in the $j^{th}$ sea and no complex strings is a 
highest-weight state with highest weight $\mu$ given by}
\be
\mu = \sum_{j=1}^{\n-1} \mu^{j} \nu^{(j)} \,,
\ee
{\it where $\mu^{j}$ are the fundamental weights of $SU(\n)$} (see Eq. 
(\ref{important})). In the corresponding Young tableau (see Figure 
\ref{fig2}),
\begin{figure}[htb]
\centering
\begin{picture}(18,3.4)(0,2.8)
\put(0,0){\line(0,1){6}}
\put(1,0){\line(0,1){6}}
\put(2,0){\line(0,1){6}}
\put(5,0){\line(0,1){6}}
\put(6,0){\line(0,1){6}}
\put(6,0){\line(0,1){6}}
\put(7,4){\line(0,1){2}}
\put(8,4){\line(0,1){2}}
\put(11,4){\line(0,1){2}}
\put(12,4){\line(0,1){2}}
\put(13,5){\line(0,1){1}}
\put(14,5){\line(0,1){1}}
\put(17,5){\line(0,1){1}}
\put(18,5){\line(0,1){1}}
\put(0,0){\line(1,0){6}}
\put(0,1){\line(1,0){6}}
\put(0,4){\line(1,0){12}}
\put(0,5){\line(1,0){18}}
\put(0,6){\line(1,0){18}}
\put(.5,3.5){\makebox(0,0)[t]{$\vdots$}}
\put(1.5,3.5){\makebox(0,0)[t]{$\vdots$}}
\put(5.5,3.5){\makebox(0,0)[t]{$\vdots$}}
\put(3.5,.7){\makebox(0,0)[t]{$\cdots$}}
\put(3.5,4.7){\makebox(0,0)[t]{$\cdots$}}
\put(3.5,5.7){\makebox(0,0)[t]{$\cdots$}}
\put(9.5,4.7){\makebox(0,0)[t]{$\cdots$}}
\put(9.5,5.7){\makebox(0,0)[t]{$\cdots$}}
\put(15.5,5.7){\makebox(0,0)[t]{$\cdots$}}
\end{picture}
\vspace{.5in}

\caption[xxx]{\parbox[t]{.7\textwidth}{%
   Young tableau corresponding to a	general	irreducible	representation 
   of $SU(\n)$}
   }
\label{fig2}
\end{figure}
\noindent 
the number of boxes in the $i^{th}$ row is equal to 
$\sum_{j=i}^{\n -1}\nu^{(j)}$. Equivalently, the representation can be 
denoted by the number of boxes in each column of the Young tableau:
\be
\left[ \underbrace{\n-1\,, \ldots \,, \n-1}_{\nu^{(\n-1)}}\,, 
       \underbrace{\n-2\,, \ldots \,, \n-2}_{\nu^{(\n-2)}}\,, \ldots \,, 
       \underbrace{1\,, \ldots \,, 1}_{\nu^{(1)}} \right] \,. \non 
\ee 
In particular, the state with a single hole in 
the $k^{th}$ sea (i.e., $\nu^{(j)} = \delta_{j,k}$)
is the highest weight of the fundamental representation $[k]$, corresponding 
to the Young tableau shown in Figure \ref{fig1}.

We label the holes in the range (\ref{range}) by 
$\{ \tilde J_{\alpha}^{(1,j)} \} \,, \alpha = 1 \,, \ldots \,, \nu^{(j)}$. 
The corresponding hole rapidities $\{ \tilde\lambda_{\alpha}^{(j)} \}$ 
are defined by 
\be
h^{(1,j)}(\tilde\lambda_{\alpha}^{(j)}) =  \tilde J_{\alpha}^{(1,j)} 
\,, \qquad \alpha = 1 \,, \ldots \,, \nu^{(j)} \,,
\ee
where $h^{(1,j)}(\lambda)$ is the counting function given in Eq.  
(\ref{hn}) with $n=1$.

For $N \rightarrow \infty$ and for each value of $j$, the roots 
$\{ \lambda_\alpha^{(1,j)}\}$ become dense on the real line, and are 
described by the corresponding densities $\sigma^{(j)}(\lambda)$ given by
\be
\sigma^{(j)}(\lambda) = {1\over N} {d \over d\lambda} h^{(1,j)}(\lambda)
\,.
\label{definesigma}
\ee
Approximating the sums in $h^{(1,j)}(\lambda)$ by integrals 
using \footnote{Here $g(\lambda)$ 
is an arbitrary function of $\lambda$ which goes to $0$ for 
$\lambda \rightarrow \pm\infty$.}
\be
{1\over N} \sum_{\alpha=1}^{M^{(1,j)}} g(\lambda_\alpha^{(1,j)}) \approx 
\int_{-\infty}^\infty  g(\lambda')\ \sigma^{(j)}(\lambda')\ d\lambda' 
- {1\over N} \sum_{\alpha=1}^{\nu^{(j)}} g(\tilde\lambda_\alpha^{(j)}) 
\label{approximation} 
\ee 
leads to a system of linear integral equations
\be
\sum_{m=1}^{\n-1}\left( \left( \delta + {\cal K} \right)_{jm} * 
\sigma^{(m)}\right) (\lambda) &=& a_{1}(\lambda) \delta_{j,1} 
+ {1\over N} \sum_{m=1}^{\n-1} \sum_{\alpha=1}^{\nu^{(m)}} 
{\cal K}(\lambda - \tilde\lambda_\alpha^{(m)})_{jm} \,, \non  \\ 
& & \qquad  \qquad \qquad \qquad j = 1\,, \ldots \,, \n -1 \,, 
\ee
where 
\be
{\cal K}(\lambda)_{jm} 
&=& a_{2}(\lambda) \delta_{m,j} 
- a_{1}(\lambda) (\delta_{m,j-1} + \delta_{m,j+1}) \,, \non \\
a_n(\lambda) &=& {1\over 2\pi} {d q_n (\lambda)\over d\lambda} 
= {1\over 2\pi} {n\over \lambda^2 + {n^2\over 4}}  \,, 
\label{KK}
\ee
as usual $*$ denotes the convolution
\be
\left( f * g \right) (\lambda) = \int_{-\infty}^\infty
f(\lambda - \lambda')\ g(\lambda')\ d\lambda' \,, 
\ee 
and $\nu^{(0)} = \nu^{(\n)} = 0$. 

This system of equations is solved by Fourier transforms, for which we 
use the following conventions:
\be
\hat f(\omega) \equiv \int_{-\infty}^\infty e^{i \omega \lambda}\ 
f(\lambda)\ d\lambda \,, \qquad\qquad
f(\lambda) = {1\over 2\pi} \int_{-\infty}^\infty e^{-i \omega \lambda}\ 
\hat f(\omega)\ d\omega \,,
\ee 
and therefore
\be
\hat a_n (\omega) = e^{-n |\omega|/2}  \,, \qquad n>0 \,.
\ee
The resolvent
\be
{\cal R}_{m m'}(\lambda) = \left( \delta(\lambda) + 
{\cal K}(\lambda)\right)^{-1}_{m m'}
\ee
has the Fourier transform \cite{sutherland}
\be
\hat {\cal R}_{m m'}(\omega) = {e^{|\omega|/2} 
\sinh \left( m_{<} |\omega|/2 \right) 
\sinh \left( (\n - m_{>})|\omega|/2 \right) \over
\sinh \left( \n |\omega|/2 \right) 
\sinh \left( |\omega|/2 \right)} 
\,,
\label{inverse}
\ee 
where $m_{>}=\max(m \,, m')$ and $m_{<}=\min(m \,, m')$.
The densities $\sigma^{(j)}(\lambda)$ are therefore given by
\be
\sigma^{(j)}(\lambda) = s^{(j)}(\lambda) 
+ {1\over N} \sum_{m=1}^{\n-1} \sum_{\alpha=1}^{\nu^{(m)}}
\left[ \delta(\lambda - \tilde\lambda_{\alpha}^{(m)}) \delta_{j,m}
- {\cal R}_{jm}(\lambda - \tilde\lambda_{\alpha}^{(m)}) \right] \,,
\label{densities/nostrings}
\ee
where the ground state densities $s^{(j)}(\lambda)$ have the Fourier 
transforms
\be
\hat s^{(j)}(\omega) = {\sinh \left( (\n - j)|\omega|/2 \right)\over
\sinh \left(\n |\omega|/2 \right)} \,.
\label{groundstate}
\ee
Heuristically, the $1/N$ terms in Eq.  (\ref{densities/nostrings}) 
describe the ``polarization'' of the Fermi seas due to the presence of 
holes.

We remark that Eq.  (\ref{numberholes}) can be solved for the integers 
$M^{(1,j)}$ in terms of the number of holes:
\be
M^{(1,j)} = \sum_{k=1}^{\n-1} \hat {\cal R}_{jk}(0) 
\left( - \nu^{(k)} + N \delta_{k,1} \right) \,.
\ee
In particular, the state with a single hole in the $k^{th}$ sea lies 
in the sector where $(N-k)/\n$ is an integer.  This is a 
generalization of the fact \cite{FT}, \cite{DMN} that for the 
Heisenberg chain ($\n=2$), the state with one hole lies in the sector 
$N=$ odd.

The energy and momentum are given by 
\be
E &=& N e_{0} + \pi \sum_{j=1}^{\n-1} \sum_{\alpha=1}^{\nu^{(j)}} 
s^{(j)}(\tilde \lambda_{\alpha}^{(j)}) \,, \non \\
P &=& N p_{0} + \sum_{j=1}^{\n-1} \sum_{\alpha=1}^{\nu^{(j)}} 
p^{(j)}(\tilde \lambda_{\alpha}^{(j)}) \,, 
\label{additive}
\ee
where the ground state energy and momentum per site are given by
\be
e_{0} = -{1\over \n}\left[ \psi(1) - \psi({1\over \n}) \right] 
\,, \qquad 
p_{0} = \pi \left({\n - 1\over \n} \right) \,,
\ee
where $\psi(z) = {d\over dz} \log \Gamma(z)$, and $p^{(j)}(\lambda)$ 
satisfies
\be
{d\over d\lambda}p^{(j)}(\lambda) = 2 \pi s^{(j)}(\lambda) \,,
\qquad p^{(j)}(0) = - \pi \left({\n - j\over \n} \right) \,. 
\label{needlater}
\ee
In particular, a single hole in the $k^{th}$ sea (which, as we have seen 
above, is in the fundamental representation $[k]$ of $SU(\n)$) with 
rapidity $\tilde\lambda^{(k)}$ is a particle-like excitation with 
energy $\pi s^{(k)}(\tilde\lambda^{(k)})$ and momentum 
$p^{(k)}(\tilde\lambda^{(k)})$.

\vspace{.2in}
\noindent
{\bf Case b: including strings of length 2}
\vspace{.2in}

\noindent 
In section 2.4, we compute the full two-particle scattering 
matrix for the cases $[1] \otimes [1]$ and $[1] \otimes [\n-1]$.  Each 
of these tensor products decomposes into a direct sum of two 
irreducible representations $[1] \otimes [1] = [1 \,, 1] \oplus [2]$ 
and $[1] \otimes [\n-1] = [\n-1 \,, 1] \oplus [\n]$, 
\begin{figure}[htb]
\centering
\be 
\begin{picture}(1,1)(0,.25)
\put(0,0){\line(0,1){1}}
\put(1,0){\line(0,1){1}}
\put(0,0){\line(1,0){1}}
\put(0,1){\line(1,0){1}}
\end{picture}
\otimes
\begin{picture}(1,1)(0,.25)
\put(0,0){\line(0,1){1}}
\put(1,0){\line(0,1){1}}
\put(0,0){\line(1,0){1}}
\put(0,1){\line(1,0){1}}
\end{picture}
=
\begin{picture}(2,1)(0,.25)
\put(0,0){\line(0,1){1}}
\put(1,0){\line(0,1){1}}
\put(2,0){\line(0,1){1}}
\put(0,0){\line(1,0){2}}
\put(0,1){\line(1,0){2}}
\end{picture}
\; \oplus
\begin{picture}(2,2)(0,.75)
\put(0,0){\line(0,1){2}}
\put(1,0){\line(0,1){2}}
\put(0,0){\line(1,0){1}}
\put(0,1){\line(1,0){1}}
\put(0,2){\line(1,0){1}}
\end{picture} \non 
\ee
\caption[xxx]{\parbox[t]{.7\textwidth}{%
 	   Young tableaux corresponding to $[1] \otimes [1] = 
 	   [1 \,, 1] \oplus [2]$}
 	   }
 	   \label{fig3}
\end{figure}
\begin{figure}[htb]
\centering
\be 
\begin{picture}(1,1)(0,.25)
\put(0,0){\line(0,1){1}}
\put(1,0){\line(0,1){1}}
\put(0,0){\line(1,0){1}}
\put(0,1){\line(1,0){1}}
\end{picture}
\otimes
\begin{picture}(1,3.4)(0,2.8)
\put(0,0){\line(0,1){6}}
\put(1,0){\line(0,1){6}}
\put(0,0){\line(1,0){1}}
\put(0,1){\line(1,0){1}}
\put(0,4){\line(1,0){1}}
\put(0,5){\line(1,0){1}}
\put(0,6){\line(1,0){1}}
\put(.5,3.5){\makebox(0,0)[t]{$\vdots$}}
\end{picture}
=
\begin{picture}(2,3.4)(0,2.8)
\put(0,0){\line(0,1){6}}
\put(1,0){\line(0,1){6}}
\put(2,5){\line(0,1){1}}
\put(0,0){\line(1,0){1}}
\put(0,1){\line(1,0){1}}
\put(0,4){\line(1,0){1}}
\put(0,5){\line(1,0){2}}
\put(0,6){\line(1,0){2}}
\put(.5,3.5){\makebox(0,0)[t]{$\vdots$}}
\end{picture}
\oplus \;
\begin{picture}(1,3.4)(0,3.2)
\put(0,0){\line(0,1){7}}
\put(1,0){\line(0,1){7}}
\put(0,0){\line(1,0){1}}
\put(0,1){\line(1,0){1}}
\put(0,4){\line(1,0){1}}
\put(0,5){\line(1,0){1}}
\put(0,6){\line(1,0){1}}
\put(0,7){\line(1,0){1}}
\put(.5,3.5){\makebox(0,0)[t]{$\vdots$}}
\end{picture} \non 
\ee
\vspace{.3in}
\caption[xxx]{\parbox[t]{.7\textwidth}{%
 	   Young tableaux corresponding to $[1] \otimes [\n-1] = 
 	   [\n-1 \,, 1] \oplus [\n]$}
 	   }
 	   \label{fig4}
\end{figure}
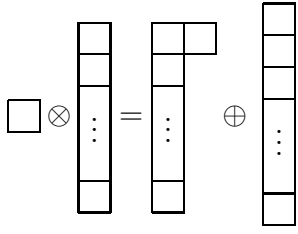
in accordance with the Young tableaux in Figures \ref{fig3} and 
\ref{fig4}, respectively. The calculation of the $S$ matrix 
eigenvalues requires the densities of the two-particle states
corresponding to the irreducible representations.
The two-particle Bethe Ansatz states which are highest weights of 
$[1 \,, 1]$ and $[\n-1 \,, 1]$ are of the form described above, with 
only real roots and two holes. Indeed, these states have
$\nu^{(j)}= 2 \delta_{j,1}$ and $\nu^{(j)}= \delta_{j,1} + \delta_{j,\n-1}$,
respectively. However, the two-particle states which 
are highest weights of $[2]$ and $[\n]$ are {\it not} of the above 
form.  In order to describe these states, we need in addition 
to two holes also strings of length 2.

Let us consider the more general case of Bethe Ansatz states with 
arbitrary values of $\nu^{(j)}$ and $M^{(2,j)}$, with $M^{(n,j)}=0$ 
for $n>2$.  We refer to this as case {\bf b}.  For this case, we 
obtain the following generalization of our previous result 
(\ref{nice}) for the eigenvalues of the Cartan generators:
\be
S^{(j)} = \nu^{(j)} + M^{(2,j-1)} + M^{(2,j+1 )} - 2 M^{(2,j)}\,.
\label{nice2}
\ee
Moreover, in the expression for $J_{\max}^{(2,j)}$ we can eliminate 
the dependence on $\{ M^{(1,k)} \}$:
\be
J_{\max}^{(2,j)} = {1\over 2}\left[
\nu^{(j)} + M^{(2,j-1)} + M^{(2,j+1 )} - M^{(2,j)} - 1 \right] \,.
\label{jmax2}
\ee
The densities are now given by
\be
\sigma^{(j)}_{{\bf b}}(\lambda) = \sigma^{(j)}_{{\bf a}}(\lambda)
- {1\over N}\sum_{\alpha=1}^{M^{(2,j)}} 
a_{1}(\lambda - \lambda_{\alpha}^{(2,j)}) \,, \qquad 
j = 1\,, \ldots \,, \n-1 \,, 
\label{densities/2strings}
\ee
where $\sigma^{(j)}_{{\bf a}}(\lambda)$ is the density 
given in Eq.  (\ref{densities/nostrings}) for the corresponding state 
with the same holes but no complex strings.  In order to determine the 
centers $\lambda_{\alpha}^{(2,j)}$ of the 2-strings in terms of the 
hole positions, we recall that $h^{(2,j)}(\lambda_{\alpha}^{(2,j)}) 
= J_{\alpha}^{(2,j)}$.  Passing from sums to integrals, we obtain the 
relations
\be
2 \pi J_{\alpha}^{(2,j)} &=& 
\sum_{m=1}^{\n -1}\sum_{\beta=1}^{M^{(2,m)}} \left[
- q_{2}(\lambda_{\alpha}^{(2,j)} - \lambda_{\beta}^{(2,m)}) \delta_{m,j}
+ q_{1}(\lambda_{\alpha}^{(2,j)} - \lambda_{\beta}^{(2,m)}) 
\left( \delta_{m,j-1} + \delta_{m,j+1} \right) \right] \non \\
& & + \sum_{\beta=1}^{\nu^{(j)}} 
q_{1}(\lambda_{\alpha}^{(2,j)} - \tilde\lambda_{\beta}^{(j)}) 
\,, \qquad \alpha = 1\,, \ldots \,, M^{(2,j)} \,, \quad 
j = 1\,, \ldots \,, \n-1 \,. 
\label{relations}
\ee
The energy and momentum are given by the same expressions in Eq.  
(\ref{additive}).

We now specialize to the cases of interest.  We see from Eq.  
(\ref{nice2}) that the two-particle Bethe Ansatz state which is the 
highest weight of $[2]$ has two holes in sea 1 and one 2-string in sea 
1 (i.e., $\nu^{(j)}= 2 \delta_{j,1}$ and $M^{(2,j)}= \delta_{j,1}$).  
Eq.  (\ref{jmax2}) implies that $J_{\max}^{(2,1)} = 0$ and hence 
$J_{1}^{(2,1)} = 0$.  From the first relation in Eq.  
(\ref{relations}) we conclude that the center $\lambda_{1}^{(2,1)}$ of 
the 2-string is midway between the two holes
\be
\lambda_{1}^{(2,1)} = {1\over 2}\left( \tilde\lambda_{1}^{(1)} + 
\tilde\lambda_{2}^{(1)} \right) \,,
\label{center1}
\ee
independently of the value of $\n$.

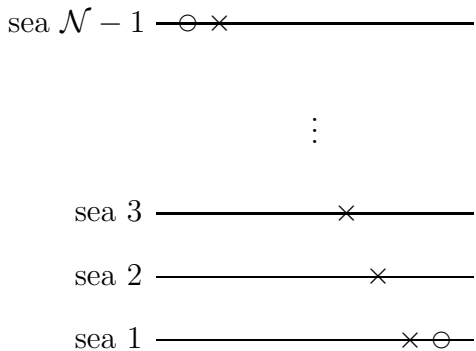
\begin{figure}[htb]
\vspace{.2in}
\centering
\begin{picture}(10,10)
\put(0,0){\line(1,0){10}}
\put(0,2){\line(1,0){10}}
\put(0,4){\line(1,0){10}}
\put(0,10){\line(1,0){10}}
\put(1,10){\circle{.5}}
\put(9,0){\circle{.5}}
\put(5,7.5){\makebox(0,0)[t]{$\vdots$}}
\put(-1.5,.4){\makebox(0,0)[t]{sea 1}}
\put(-1.5,2.4){\makebox(0,0)[t]{sea 2}}
\put(-1.5,4.4){\makebox(0,0)[t]{sea 3}}
\put(-2.5,10.4){\makebox(0,0)[t]{sea $\n-1$}}
\put(2,10.35){\makebox(0,0)[t]{$\times$}}
\put(7,2.35){\makebox(0,0)[t]{$\times$}}
\put(6,4.35){\makebox(0,0)[t]{$\times$}}
\put(8,.35){\makebox(0,0)[t]{$\times$}}
\end{picture}
\vspace{.2in}
\caption[xxx]{\parbox[t]{.7\textwidth}{
	The singlet $[\n]$ two-particle Bethe Ansatz state. Horizontal lines 
	represent the $\n-1$ root distributions forming the Fermi seas, circles 
	denote holes in  these seas, and crosses mark the centers of the 2-strings.}
	}
	\label{fig5}
\end{figure}

Finally, we consider the singlet $[\n]$ two-particle Bethe Ansatz 
state.  This is the state with one hole in sea 1, one hole in sea 
$\n-1$, and one 2-string in {\it each} of the $\n-1$ seas (i.e., 
$\nu^{(j)}= \delta_{j,1} + \delta_{j,\n-1}$ and $M^{(2,j)}= 1$ for $j 
= 1\,, \ldots \,, \n-1$).  We observe that each $J_{1}^{(2,j)} = 0$.  
Remarkably, the relations (\ref{relations}) lead to a linear system of 
equations for $\lambda_{1}^{(2,j)}$ whose resolvent is $\hat 
{\cal R}_{jk}(0)$.  We obtain for the centers of the 2-strings the result
\be
\lambda_{1}^{(2,j)} = \tilde\lambda_{1}^{(1)} 
+ {j\over \n}\left( \tilde\lambda_{1}^{(\n-1)} - \tilde\lambda_{1}^{(1)} 
\right) \,, \qquad j = 1\,, \ldots \,, \n-1 \,. 
\label{center2}
\ee 
This result is represented schematically (for 
$\tilde\lambda_{1}^{(1)} > \tilde\lambda_{1}^{(\n -1)}$) in Figure \ref{fig5}.

\subsection{Bulk $S$ matrix}

Following Refs.  \cite{korepin}, \cite{andrei/destri}, we define the 
two-particle $S$ matrix $S^{[j]\otimes[k]}$ for particles of type 
$[j]$ and $[k]$ by the momentum quantization condition
\be
\left( e^{ip^{(j)}(\tilde\lambda^{(j)}) N} 
S^{[j]\otimes[k]} - 1 \right) 
|\tilde\lambda^{(j)}\,, \tilde\lambda^{(k)} \rangle = 0 \,,
\label{quantization}
\ee 
where the single-particle momentum $p^{(j)}(\lambda)$ is given 
by Eq. (\ref{needlater}), and $\tilde\lambda^{(j)}\,, \tilde\lambda^{(k)}$ 
are the corresponding hole rapidities.

We focus our attention on the cases $[1] \otimes [1]$ and $[1] 
\otimes [\n-1]$, for which the $S$ matrices act in $C_{\n} \otimes 
C_{\n}$ and $C_{\n} \otimes \overline{C_{\n}}$, respectively.  
\footnote{The case $[\n-1] \otimes [\n-1]$ is equivalent to the case
$[1] \otimes [1]$.  Other cases can presumably be treated along 
similar lines; however, they involve higher-dimensional 
representations of $SU(\n)$, and the corresponding $S$ matrices have 
more than two distinct eigenvalues.}
As already noted, for these cases, the tensor product $[j] \otimes 
[k]$ decomposes into a direct sum of precisely two irreducible 
representations. (See Figures \ref{fig3} and \ref{fig4}.)
The Bethe Ansatz states which are highest weights of 
these irreducible representations belong to the cases we denoted {\bf a} 
and {\bf b}, with densities given by Eqs. (\ref{densities/nostrings}) and 
(\ref{densities/2strings}), respectively.
Specifically, the state corresponding to the completely antisymmetric 
Young tableau (i.e., with only 1 column) belongs to case {\bf b}, and 
the other state belongs to case {\bf a}.

We now compute the eigenvalues of $S^{[j]\otimes[k]}$.  Let $S_{\bf 
a}$ and $S_{{\bf b}}$ be the eigenvalues of $S^{[j]\otimes[k]}$ 
corresponding to states belonging to cases {\bf a} and {\bf b}, 
respectively.  The identity
\be 
{1\over 2\pi} {d\over d\lambda}p^{(j)}(\lambda) 
+ \sigma^{(j)}(\lambda) - s^{(j)}(\lambda)  = 
{1\over N} {d\over d\lambda}h^{(1,j)}(\lambda) 
\label{identity}
\ee
can easily be obtained from Eqs.  (\ref{definesigma}) and 
(\ref{needlater}).  Integrating from $-\infty$ to 
$\tilde\lambda^{(j)}$ and exponentiating, we obtain the relation
\be
e^{i p^{(j)}(\tilde\lambda^{(j)}) N}\ 
e^{i 2\pi N \int_{-\infty}^{\tilde\lambda^{(j)}}
\left( \sigma^{(j)}(\lambda) - s^{(j)}(\lambda) \right)\ d\lambda}\
e^{i 2\pi\left( h^{(1,j)}(-\infty) - \tilde J^{(1,j)}\right)}\
e^{-i N p^{(j)}(-\infty)} = 1 \,,
\label{desired}
\ee
where $h^{(1,j)}(\tilde\lambda^{(j)}) = \tilde J^{(1,j)}$.
Comparing with Eq. (\ref{quantization}), we see that (up to a 
rapidity-independent phase factor)
\be
S_{{\bf a}} \sim   
\exp \left\{ i 2\pi N \int_{-\infty}^{\tilde\lambda^{(j)}}
\left( \sigma_{{\bf a}}^{(j)}(\lambda) - s^{(j)}(\lambda) \right) d\lambda
\right\}
\,,
\label{sim}
\ee 
where $\sigma_{{\bf a}}^{(j)}(\lambda)$ is given by Eq.  
(\ref{densities/nostrings}) with $\nu^{(m)} = \delta_{m,j} + 
\delta_{m,k}$.  The integral can be explicitly performed using the 
Fourier-space expression (\ref{inverse}), as well as the identity 
\cite{GR}
\be
\int_{0}^{\infty}{\left( 1 - e^{-\beta x} \right) 
\left( 1 - e^{-\gamma x} \right) e^{-\mu x}\over 1 - e^{-x}}  
{dx\over x} = \log {\Gamma(\mu)\Gamma(\mu + \beta + \gamma)\over 
\Gamma(\mu + \beta)\Gamma(\mu + \gamma)} \,,
\label{GRintegral}
\ee
provided $Re\ \mu >0$, $Re\ \mu > - Re\ \beta$, $Re\ \mu > - Re\ \gamma$, 
and $Re\ \mu > -Re\ (\beta + \gamma)$. One finds \cite{kulish/reshetikhin}
\be
S_{{\bf a}} &=& \prod_{l=0}^{j-1}
{\Gamma\left( 1 + {2l - j - k\over 2\n} - {i \tilde\lambda\over \n} \right)
\Gamma\left( {2l + k - j\over 2\n}  + {i \tilde\lambda\over \n} \right) \over
\Gamma\left( 1 + {2l - j - k\over 2\n} + {i \tilde\lambda\over \n} \right)
\Gamma\left( {2l + k - j\over 2\n}  - {i \tilde\lambda\over \n} \right) }
\,, \qquad 
\tilde\lambda \equiv  \tilde\lambda^{(j)} - \tilde\lambda^{(k)} \,.
\ee 
In particular, for the states $[1 \,, 1]$ and $[\n -1 \,, 1]$,
\be
S_{[1 \,, 1]} &=& {\Gamma\left( 1 - {1\over \n}(1 + i \tilde\lambda)\right)
\Gamma\left( 1 + {i \tilde\lambda\over \n}\right)\over
\Gamma\left( 1 + {1\over \n}(-1 + i \tilde\lambda)\right)
\Gamma\left( 1 - {i \tilde\lambda\over \n}\right)} \,, \qquad 
\tilde\lambda \equiv  \tilde\lambda_{1}^{(1)} - \tilde\lambda_{2}^{(1)} 
\,, \label{s1} \\
S_{[\n -1 \,, 1]} &=& 
{\Gamma\left( {1\over 2} - {i \tilde\lambda\over \n}\right)
\Gamma\left( {1\over 2} + {1\over \n}(-1 + i \tilde\lambda)\right)\over
\Gamma\left( {1\over 2} + {i \tilde\lambda\over \n}\right)
\Gamma\left( {1\over 2} - {1\over \n}(1 + i \tilde\lambda)\right)} \,, \qquad
\tilde\lambda \equiv  \tilde\lambda_{1}^{(1)} - \tilde\lambda_{1}^{(\n -1)} 
\,. \label{s2}
\ee

Although $S_{{\bf a}}$ has been determined only up to a rapidity-independent 
phase factor, the ratio $S_{{\bf b}}/S_{{\bf a}}$ can be computed exactly:
\be
{S_{{\bf b}}\over S_{{\bf a}}} &=&
e^{i 2\pi N \int_{-\infty}^{\tilde\lambda^{(j)}} \left( 
\sigma_{{\bf b}}^{(j)}(\lambda) - \sigma_{{\bf a}}^{(j)}(\lambda) \right)
d\lambda}\
e^{i 2\pi\left( h_{{\bf b}}^{(1,j)}(-\infty) 
-  h_{{\bf a}}^{(1,j)}(-\infty) \right)}\ 
e^{-i 2\pi\left( \tilde J_{{\bf b}}^{(1,j)} 
- \tilde J_{{\bf a}}^{(1,j)} \right)} \non  \\
&=& \prod_{\alpha=1}^{M^{(2,j)}}
e_1(\tilde\lambda^{(j)} - \lambda_{\alpha}^{(2,j)}) \,,
\label{ratio}
\ee
where $\sigma_{{\bf b}}^{(j)}(\lambda)$ is given by Eq. (\ref{densities/2strings}).
In particular, for the states $[2]$ and $[\n]$, we find
\be
{S_{[2]}\over S_{[1 \,, 1]}} = e_1 \left( {1\over 2}
(\tilde\lambda_{1}^{(1)} - \tilde\lambda_{2}^{(1)}) \right) \,, \\
{S_{[\n]}\over S_{[\n -1 \,, 1]}} = e_{1}\left( {1\over \n}
(\tilde\lambda_{1}^{(1)} - \tilde\lambda_{1}^{(\n -1)}) \right)
\,,
\ee
where we have used our results for the centers of the 2-strings 
(\ref{center1}) and (\ref{center2}), respectively.

Finally, we cast our results into matrix form.  For the case 
$[1]\otimes[1]$, $SU(\n)$ symmetry implies that the complete 
two-particle $S$ matrix is given by
\be
S^{[1]\otimes[1]} &=& S_{[1 \,, 1]}(\tilde \lambda)\
{1\over 2} \left( 1 + {\cal P} \right) 
+ S_{[2]}(\tilde \lambda)\ {1\over 2}\left( 1 - {\cal P} \right)  \non \\
&=& S_{[1,1]}(\tilde \lambda)\ 
{i\tilde \lambda + {\cal P}\over i\tilde \lambda + 1} \,, \qquad 
\tilde\lambda=\tilde\lambda_{1}^{(1)} - \tilde\lambda_{2}^{(1)} \,, 
\ee
where $S_{[1 \,, 1]}(\tilde \lambda)$ is given by Eq. (\ref{s1}).
Moreover, for the case $[1] \otimes [\n-1]$, 
\be
S^{[1] \otimes [\n-1]} &=& S_{[\n-1 \,, 1]}(\tilde \lambda)\ 
\left( 1 - P_{[\n]} \right)+ S_{[\n]}(\tilde \lambda)\  P_{[\n]} \non \\
&=& S_{[\n-1 \,, 1]}(\tilde \lambda)\  
\left( 1 - {\n \over i\tilde\lambda 
+ {\n\over 2}} P_{[\n]} \right) \,, \qquad 
\tilde\lambda = \tilde\lambda_{1}^{(1)} - \tilde\lambda_{1}^{(\n -1)} \,,
\ee
where $P_{[\n]}$ is the projector onto the one-dimensional subspace 
$[\n]$, namely, $P_{[\n]} = {1\over \n} {\cal P}_{12}^{t_{2}}$; and 
$S_{[\n-1 \,, 1]}(\tilde \lambda)$ is given by Eq. (\ref{s2}).
These results agree  with those found using the bootstrap approach by 
Ogievetsky {\it et al.}  \cite{ORW}\footnote{We take into account apparent 
typos in their Eqs. (2.19) and (2.23).}, without additional CDD factors.

\section{The open chain}

We turn now to the open integrable chain constructed with the 
$SU(\n)$-invariant $R$ matrix (\ref{Rmatrix}) and the diagonal $K$ 
matrices (\ref{Kmatrices}).  Although this $R$ matrix does not have 
crossing symmetry for $\n > 2$, it does have the property \cite{RSTS}
\be
\left( \left( \left( R_{12}(\lambda)^{t_{2}}\right)^{-1} \right)^{t_{2}} 
\right)^{-1} \propto  M_{2}\ R_{12}(\lambda + 2\rho )\ M_{2}^{-1} \,,
\ee
with $M=1$ and $\rho = \n/2$.
One can therefore prove \cite{sklyanin}, \cite{mn/addendum} the 
commutativity of the transfer matrix $t_{(l)}(\lambda \,, \xi_{\mp})$ given by 
\footnote{The more general transfer matrix 
$t_{(l_{+}, l_{-})}(\lambda\,, \xi_{\mp})$ constructed with 
$K^{\mp}_{(l_{\mp})}(\lambda \,, \xi_{\mp})$ also forms a one-parameter 
commutative family.  For simplicity, we consider here the 
special case $l_{+} = l_{-} = l$.}
\be
t_{(l)}(\lambda\,, \xi_{\mp}) = \tr_{0} K^{+}_{(l)}{}_{0}(\lambda \,, \xi_{+})\  
T_{0}(\lambda)\  K^{-}_{(l)}{}_{0}(\lambda \,, \xi_{-})\ \hat T_{0}(\lambda)\,,
\label{transferopen}
\ee
where $T_{0}(\lambda)$ is the monodromy	matrix (\ref{monodromy}) and 
$\hat T_{0}(\lambda)$ is given by
\be
\hat T_{0}(\lambda) = R_{10}(\lambda) \cdots  R_{N0}(\lambda) \,. 
\label{hatmonodromy}
\ee
The Hamiltonian (\ref{openHamiltonian}) is related to the 
derivative of the transfer matrix at $\lambda=0$ 
\be
{\cal H}_{open} = {1 \over 4 \xi_{-} \tr K^{+}_{(l)}(0 \,, \xi_{+})}  
{d\over d \lambda} t_{(l)}(\lambda\,, \xi_{\mp})
\Big\vert_{\lambda=0} - {i\over 4} 
{d\over d \lambda} \log \tr K^{+}_{(l)}(\lambda \,, \xi_{+})\Big\vert_{\lambda=0}
\,.
\ee

\subsection{Symmetries of the transfer matrix}

The $SU(\n)$ invariance (\ref{invariance}) of the $R$ matrix implies 
that
\be
U_{2}\ R_{12}(\lambda)\ U_{2}^{\dagger} =  U_{1}^{\dagger}\ R_{12}(\lambda)\ U_{1}
\ee
for all $U \in SU(\n)$. 
The LHS can be regarded as a ``quantum-space'' transformation, while 
the RHS can be regarded as an ``auxiliary-space'' transformation. 
The quantum-space operator ${\cal U}$ defined by
\be
{\cal U} = U_{1}\ U_{2} \cdots U_{N} 
\label{calU}
\ee
therefore has the following action on the monodromy matrices
\be
{\cal U}\ T_{0}(\lambda)\ {\cal U}^{\dagger} 
&=&  U_{0}^{\dagger}\ T_{0}(\lambda)\ U_{0} \,, \non \\
{\cal U}\ \hat T_{0}(\lambda)\ {\cal U}^{\dagger} 
&=&  U_{0}^{\dagger}\ \hat T_{0}(\lambda)\ U_{0} \,,
\ee
and the transfer matrix transforms as follows:
\be
{\cal U}\ t_{(l)}(\lambda\,, \xi_{\mp})\ {\cal U}^{\dagger} 
=  \tr_{0} \left\{ \left( U\ K^{+}_{(l)}(\lambda \,, \xi_{+})\  U^{\dagger} 
\right){}_{0}\
T_{0}(\lambda)\ \left( U\ K^{-}_{(l)}(\lambda \,, \xi_{-})\ 
U^{\dagger} \right){}_{0}\ \hat T_{0}(\lambda) \right\} \,.
\ee
For $U \in SU(l) \times SU(\n-l) \times U(1)$, evidently
\be
U\ K^{\mp}_{(l)}(\lambda \,, \xi_{\mp})\ U^{\dagger} =  K^{\mp}_{(l)} 
\,,
\ee
and therefore
\be
{\cal U}\ t_{(l)}(\lambda\,, \xi_{\mp})\ {\cal U}^{\dagger} 
= t_{(l)}(\lambda\,, \xi_{\mp}) \,.
\ee
That is, the transfer matrix has the invariance $SU(l) \times SU(\n-l) 
\times U(1)$.  In particular, it commutes with all the $SU(\n)$ Cartan 
generators
\be
\left[ t_{(l)}(\lambda\,, \xi_{\mp})  \,, S^{(k)} \right] = 0 \,, 
\qquad k = 1 \,, \ldots \,, \n-1 \,.
\ee 

The transfer matrix also has a less evident --- but very useful --- 
``duality'' symmetry which maps $l \leftrightarrow \n -l$.  This 
symmetry originates from the simple fact that under the 
transformations
\be
\xi_{-} & \rightarrow & -\xi_{-}  \,, \non \\
\xi_{+} & \rightarrow & -\xi_{+} + \n \,,
\ee
the elements (\ref{elements}) of the $K$ matrices transform into each 
other: $a^{\mp} \leftrightarrow -b^{\mp}$. Therefore,
\be
K^{\mp}_{(l)}(\lambda \,, \xi_{\mp}) \rightarrow 
- K'^{\mp}_{(l)}(\lambda \,, \xi_{\mp}) \,,
\ee
where
\be
K'^{\mp}_{(l)}(\lambda \,, \xi_{\mp}) = diag \Bigl(  
\underbrace{b^{\mp}\,, \ldots \,, b^{\mp}}_{l} \,, 
\underbrace{a^{\mp}\,, \ldots \,, a^{\mp}}_{\n-l} \Bigr) \,.
\ee
Notice that $K'^{\mp}_{(\n - l)}$ and $K^{\mp}_{(l)}$ are related by a 
cyclic permutation. Thus, there exist matrices $U_{(l)} \in SU(\n)$ 
such that \footnote{Evidently, the $\n \times \n$ matrices $U_{(j,k)}$ 
($j \,, k = 1 \,, \ldots \,, \n$ with $j \ne k$) corresponding to rotations 
by $\pi/2$ in the $(j,k)$ plane are elements of $SU(\n)$. These 
matrices have the property of permuting the $j^{th}$ and $k^{th}$ 
elements of an arbitrary diagonal matrix $D = diag \left( d_{1} \,, 
\ldots \,, d_{\n} \right)$:
\be
U_{(j,k)}\ diag   \left( d_{1} \,, \ldots \,, d_{j} \,, \ldots \,, 
d_{k} \,, \ldots \,,  d_{\n} \right) U_{(j,k)}^{\dagger} =
diag \left( d_{1} \,, \ldots \,, d_{k} \,, \ldots \,, 
d_{j} \,, \ldots \,, d_{\n} \right) \,.
\ee 
We choose the matrices $U_{(l)}$ to be products of matrices of the type 
$U_{(j,k)}$. Note that Eq. (\ref{permute}) does not uniquely determine
$U_{(l)}$.}
\be
U_{(l)}\ K^{\mp}_{(l)}(\lambda \,, \xi_{\mp})\ U_{(l)}^{\dagger} 
=  K'^{\mp}_{(\n -l)}(\lambda \,, \xi_{\mp})
= - K^{\mp}_{(l')}(\lambda \,, \xi'_{\mp}) \,,
\label{permute}
\ee
where
\be
\xi'_{-} &=& -\xi_{-}  \,, \non \\
\xi'_{+} &=& -\xi_{+} + \n \,, \non \\ 
l' &=& \n -l \,.
\label{primes}
\ee
Correspondingly, we obtain the desired ``duality'' transformation 
property of the transfer matrix
\be
{\cal U}_{(l)}\ t_{(l)}(\lambda\,, \xi_{\mp})\ {\cal U}_{(l)}^{\dagger} 
= t_{(l')}(\lambda\,, \xi'_{\mp}) \,,
\ee
where
\be
{\cal U}_{(l)} = U_{(l)\ 1}\ U_{(l)\ 2} \cdots U_{(l)\ N} \,. \non 
\ee 
Notice that the square of this transformation is the identity.  
The transfer matrix is ``self-dual'' for $\xi_{-} = \xi'_{-} = 0$,
$\xi_{+} = \xi'_{+} = \n/2$, and $l = l' = \n/2$.

\subsection{Dual pseudovacuum}

The algebraic Bethe Ansatz is usually implemented with the pseudovacuum 
$\omega_{(1)}$, where 
\be
\omega_{(k)} = \left( \begin{array}{c}
                       0\\
                       \vdots \\
                       1 \\
                       \vdots \\
                       0
                       \end{array} \right)^{\otimes N} 
                       \vphantom{\begin{array}{c}
                       0\\
                       \vdots \\
                       1 \\
                       \end{array}} \leftarrow k^{th} \,.
\label{pseudovack} 
\ee
We shall also make use of Bethe Ansatz states constructed with the 
``dual'' pseudovacuum\footnote{We use here the matrix $U_{(l)}$ 
described in the previous footnote. A different choice of $U_{(l)}$ can 
lead to a different ``dual'' pseudovacuum. Nevertheless, 
the results given in Section 3.4 for the boundary $S$ matrices do not 
depend on this choice.}
\be
\omega' = {\cal U}_{(l)}\ \omega_{(1)} = \omega_{(\n)}
\label{dualvac}
\ee 
in order to compute boundary $S$ matrices.  We can obtain the 
corresponding Bethe Ansatz equations (BAE) with the help of the 
following

\begin{lemma}
Let $g$ be a transformation on the parameters $l$ and $\xi_{\mp}$ of 
the boundary $K$ matrices,
\be
g:\left\{ \begin{array}{c}
         l  \rightarrow  l' \\
         \xi_{\mp}  \rightarrow   \xi'_{\mp}
          \end{array} \right. \,, 
\ee 
which squares to the identity, i.e., $g^{2}=1$. 
Furthermore, let ${\cal U}_{(l)}$ be a unitary transformation on 
$C_{\n}^{\otimes N}$ such that
\be
{\cal U}_{(l)}\ t_{(l)}(\lambda\,, \xi_{\mp})\ {\cal U}_{(l)}^{\dagger} &=& 
t_{(l')}(\lambda\,, \xi'_{\mp}) \label{p1} \\
{\cal U}_{(l)}\ \omega  &=& \omega' \,. \label{p2}
\ee
Then the BAE for the transfer matrix 
$t_{(l)}(\lambda\,, \xi_{\mp})$ with the pseudovacuum $\omega'$ are 
the same as the BAE for the transfer matrix 
$t_{(l')}(\lambda\,, \xi'_{\mp})$ with the pseudovacuum $\omega$.
\end{lemma}

Before giving our general proof, it is instructive to examine a more
explicit proof for the special case $\n=2$, which was first considered 
by Sklyanin \cite{sklyanin}.  For this case $l=l'=1$, and we 
therefore suppress the label $l$ and write the transfer matrix as 
$t(\lambda\,, \xi_{\mp})$.  Moreover, we take ${\cal U}$ as in Eq.  
(\ref{calU}), with
\be
U = \left( \begin{array}{cc}
                0 & -1  \\
                1 & 0 
            \end{array} \right) \,.
\ee
Then condition (\ref{p1}) is satisfied, with $\xi'_{\mp}$ as in Eq.  
(\ref{primes}).  We consider the pseudovacuum $\omega = \omega_{(1)}$, 
and thus $\omega' = {\cal U}\ \omega_{(1)} = \omega_{(2)}$.  The 
algebraic Bethe Ansatz leads to the fundamental result
\be
t(\lambda\,, \xi_{\mp})\ | \{ \lambda_{\alpha} \} \,, \xi_{-} \rangle =
\Lambda \left( \lambda \,,  \{ \lambda_{\alpha} \} \,, \xi_{\mp} 
\right) \ | \{ \lambda_{\alpha} \} \,, \xi_{-} \rangle \,,
\label{start}
\ee
where 
\be
| \{ \lambda_{\alpha} \} \,, \xi_{-} \rangle 
= {\cal B}(\lambda_{1} \,, \xi_{-}) \cdots {\cal B}(\lambda_{M} \,, 
\xi_{-})\ \omega \,,
\ee
and $\{ \lambda_{\alpha} \}$ are solutions of the BAE with 
pseudovacuum $\omega$.
Multiplying both sides of Eq. (\ref{start}) with ${\cal U}$, and using 
condition (\ref{p1}) as well as the fact 
${\cal U}\ {\cal B}(\lambda \,, \xi_{-})\  {\cal U}^{\dagger} 
= {\cal C}(\lambda \,, \xi'_{-})$, we see that
\be
t(\lambda\,, \xi'_{\mp})\ | \{ \lambda_{\alpha} \} \,, \xi'_{-} \rangle' =
\Lambda \left( \lambda \,,  \{ \lambda_{\alpha} \} \,, \xi_{\mp} 
\right) \ | \{ \lambda_{\alpha} \} \,, \xi'_{-} \rangle' \,,
\label{middle}
\ee
where
\be
| \{ \lambda_{\alpha} \} \,, \xi_{-} \rangle' 
= {\cal C}(\lambda_{1} \,, \xi_{-}) \cdots {\cal C}(\lambda_{M} \,, 
\xi_{-})\ \omega' \,.
\ee
Changing $\xi_{\mp} \rightarrow \xi'_{\mp}$ in Eq. (\ref{middle}), we obtain
\be 
t(\lambda\,, \xi_{\mp})\ | \{ \lambda'_{\alpha} \} \,, \xi_{-} \rangle' =
\Lambda \left( \lambda \,,  \{ \lambda'_{\alpha} \} \,, \xi'_{\mp} 
\right) \ | \{ \lambda'_{\alpha} \} \,, \xi_{-} \rangle' \,,
\ee
where $\{ \lambda'_{\alpha} \}$ satisfy the same BAE as $\{ \lambda_{\alpha} 
\}$, except with $\xi_{\mp} \rightarrow \xi'_{\mp}$. 

We consider now the general case.  The nested algebraic Bethe Ansatz 
leads to the result
\be
t_{(l)}(\lambda\,, \xi_{\mp})\ | \ \rangle =
\Lambda_{(l)} \left( \lambda \,,  \xi_{\mp} \right) \ | \  \rangle \,,
\ee
where the eigenstate $| \  \rangle$ is constructed with the pseudovacuum 
$\omega$. Multiplying both sides by ${\cal U}_{(l)}$ and using 
condition (\ref{p1}) gives
\be
t_{(l')}(\lambda\,, \xi'_{\mp})\ {\cal U}_{(l)}\ | \ \rangle =
\Lambda_{(l)} \left( \lambda \,,  \xi_{\mp} \right) \ 
{\cal U}_{(l)}\ | \  \rangle \,.
\ee
Changing $\xi_{\mp} \rightarrow \xi'_{\mp}$ and $l \rightarrow l'$, 
we obtain
\be
t_{(l)}(\lambda\,, \xi_{\mp})\  | \ \rangle' =
\Lambda_{(l')} \left( \lambda \,,  \xi'_{\mp} \right) \ | \  \rangle' \,,
\ee
where $| \ \rangle' $ is constructed with the pseudovacuum $\omega' = 
{\cal U}_{(l)}\ \omega$.  Notice that the eigenvalue of the transfer matrix 
is $\Lambda_{(l')} (\lambda \,, \xi'_{\mp})$.  Recalling 
(see, e.g.,  \cite{QISM}) that the BAE are precisely the conditions that 
the eigenvalues have vanishing residues, we conclude that the BAE 
corresponding to the pseudovacuum $\omega'$ are the same as the BAE 
corresponding to the pseudovacuum $\omega$, except with $\xi_{\mp} 
\rightarrow \xi'_{\mp}$ and $l \rightarrow l'$.  This concludes our 
proof of the Lemma.

\subsection{Bethe Ansatz and multihole states}

The eigenstates of the transfer matrix $t_{(l)}(\lambda \,, \xi_{\mp})$ 
with the pseudovacuum $\omega_{(1)}$ have been constructed in 
Refs. \cite{devega/gonzalezruiz1}, \cite{devega/gonzalezruiz3}. 
The corresponding Bethe Ansatz equations (BAE) are given 
by \footnote{Starting from Eq. (17) in 
\cite{devega/gonzalezruiz1}, we make a shift of variables 
$\mu^{(k)}_{j} \rightarrow \mu^{(k)}_{j} - {k\gamma\over 2}$; 
and we then take the isotropic limit by making the redefinitions
$\gamma = i \eta$, $\xi_{\mp} \rightarrow i \eta \xi_{\mp}$, 
$\mu^{(k)}_{j} = \eta \lambda^{(k)}_{j}$, and then letting
$\eta \rightarrow 0$.}
\be
1 &=& \left[ e_{2\xi_{-} + l} (\lambda_{\alpha}^{(l)})\ 
e_{-\left( 2\xi_{+} - 2 \n + l \right)}(\lambda_{\alpha}^{(l)})\  
\delta_{l,k} + \left( 1 - \delta_{l,k} \right) \right]  \non \\  
& & \times \prod_{\beta=1}^{M^{(k-1)}} 
e_{-1}(\lambda_{\alpha}^{(k)} - \lambda_{\beta}^{(k-1)})\ 
e_{-1}(\lambda_{\alpha}^{(k)} + \lambda_{\beta}^{(k-1)})
\prod_{\stackrel{\scriptstyle\beta=1}{\scriptstyle\beta \ne \alpha}}^{M^{(k)}} 
e_{2}(\lambda_{\alpha}^{(k)} - \lambda_{\beta}^{(k)})\
e_{2}(\lambda_{\alpha}^{(k)} + \lambda_{\beta}^{(k)}) \non \\ 
& & \times \prod_{\beta=1}^{M^{(k+1)}} 
e_{-1}(\lambda_{\alpha}^{(k)} - \lambda_{\beta}^{(k+1)})\
e_{-1}(\lambda_{\alpha}^{(k)} + \lambda_{\beta}^{(k+1)}) \non \\ 
& &  \qquad \qquad 
\alpha = 1 \,, \ldots \,, M^{(k)} \,, \qquad  
k = 1\,, \ldots \,, \n-1  \,.
\label{openBAE}
\ee
As before, $M^{(0)} = N \,, \quad M^{(\n)} = 0 \,, \quad 
\lambda_{\alpha}^{(0)} = \lambda_{\alpha}^{(\n)} = 0 \,.$ The 
requirement that solutions of the BAE correspond to independent Bethe 
Ansatz states leads to the restriction $\lambda_{\alpha}^{(k)} > 0$.
For later convenience, we restrict $\xi_{-} > {1\over 2}(\n -1)$,
$\xi_{+} > \n - {1\over 2}$. (See Eqs. (\ref{deltasigma}) and 
(\ref{deltasigmaprime}) below.)

Since we need to consider only one-particle states in order to compute 
boundary $S$ matrices \cite{DMN}, we restrict our attention here to 
{\it real} solutions of the BAE, i.e., no complex strings. In terms of 
the counting functions $h^{(k)}_{(l)}(\lambda)$ defined by
\be
h^{(k)}_{(l)}(\lambda) &=& {1\over 2\pi} \Bigl\{ q_{1}(\lambda) +
\left[ -q_{2\xi_{-}+l}(\lambda)  + q_{2\xi_{+} - 2\n + l}(\lambda) 
\right] \delta_{k,l} \non \\ 
&+& \sum_{\beta=1}^{M^{(k-1)}} \left[ 
q_{1} (\lambda - \lambda_{\beta}^{(k-1)}) 
+ q_{1} (\lambda + \lambda_{\beta}^{(k-1)}) \right]
- \sum_{\beta=1}^{M^{(k)}} \left[ 
q_{2} (\lambda - \lambda_{\beta}^{(k)}) 
+ q_{2} (\lambda + \lambda_{\beta}^{(k)}) \right] \non \\ 
&+& \sum_{\beta=1}^{M^{(k+1)}} \left[ 
q_{1} (\lambda - \lambda_{\beta}^{(k+1)}) 
+ q_{1} (\lambda + \lambda_{\beta}^{(k+1)}) \right]
\Bigr\} \,, 
\label{countingopen} 
\ee 
the BAE take the form 
\be
h^{(k)}_{(l)} ( \lambda_\alpha^{(k)} ) = J_\alpha^{(k)}  \,, 
\qquad \alpha = 1\,, \ldots \,, M^{(k)} \,, 
\qquad k = 1 \,, \ldots \,, \n-1 \,. 
\label{BAlogopen} 
\ee 

Although ultimately we focus on one-hole states, it is convenient to 
first consider the more general case of multihole states.  As in the 
previous section, we let $\nu^{(j)}$ denote the number of holes in the 
$j^{th}$ sea, and we define the hole rapidities
$\{ \tilde\lambda_{\alpha}^{(j)} \}$ by
\be
h^{(j)}_{(l)}(\tilde\lambda_{\alpha}^{(j)}) =  \tilde J_{\alpha}^{(j)} 
\,, \qquad \alpha = 1 \,, \ldots \,, \nu^{(j)} \,.
\ee
In the thermodynamic limit, the roots are described by 
densities $\sigma^{(j)}_{(l)}(\lambda)$ given by
\be
\sigma^{(j)}_{(l)}(\lambda) = {1\over N} {d \over d\lambda} h^{(j)}_{(l)}(\lambda)
\,.
\label{definesigmaopen}
\ee
The sums in $h^{(j)}_{(l)}(\lambda)$ can be approximated by integrals 
using (see, e.g., \cite{GMN})
\be
{1\over N} \sum_{\alpha=1}^{M^{(j)}} g(\lambda_\alpha^{(j)}) \approx 
\int_{0}^\infty  g(\lambda')\ \sigma^{(j)}_{(l)}(\lambda')\ d\lambda' 
- {1\over N} \sum_{\alpha=1}^{\nu^{(j)}} g(\tilde\lambda_\alpha^{(j)}) 
- {1\over 2N} g(0) \,.
\label{approximationopen} 
\ee
For the symmetric density  $\sigma^{(j)}_{(l)\ s}(\lambda)$ defined by
\be
\sigma^{(j)}_{(l)\ s}(\lambda) = 
\left\{ \begin{array}{cc}
           \sigma^{(j)}_{(l)}(\lambda)   & \lambda > 0 \\
           \sigma^{(j)}_{(l)}(-\lambda)  & \lambda < 0   
         \end{array} \right.
\ee   
we obtain the system of linear integral equations
\be
& &\sum_{m=1}^{\n-1} \left( \left( \delta + {\cal K} \right)_{jm} * 
\sigma^{(m)}_{(l)\ s}\right) (\lambda) = 2 a_{1}(\lambda) \delta_{j,1} 
+ {1\over N} \Big\{ a_{2}(\lambda) 
+ a_{1}(\lambda) \left( -1 + \delta_{j,1} + \delta_{j,\n -1} \right) \non \\
&+& \left( - a_{2\xi_{-} + l}(\lambda) 
+ a_{2\xi_{+} - 2 \n + l}(\lambda) \right) \delta_{j,l} 
+ \sum_{m=1}^{\n-1} \sum_{\alpha=1}^{\nu^{(m)}} 
\left( {\cal K}(\lambda - \tilde\lambda_\alpha^{(m)})_{jm} +
{\cal K}(\lambda + \tilde\lambda_\alpha^{(m)})_{jm} \right) \Big\} \,, \non \\
& & \qquad \qquad \qquad \qquad \qquad j = 1\,, \ldots \,, \n -1 \,, 
\ee
where ${\cal K}(\lambda)_{jm}$ is defined in Eq.  (\ref{KK}).  The solution is 
given by
\be
\sigma^{(j)}_{(l)\ s}(\lambda) &=& 2 s^{(j)}(\lambda) 
+ \delta \sigma^{(j)}_{(l)}(\lambda)
+ {1\over N}  \Big\{ \sum_{m=1}^{\n-1} 
\left( {\cal R}_{jm} * \left[ a_{2}
+ a_{1} \left( -1 + \delta_{m,1} + \delta_{m,\n -1} \right) \right] 
\right)(\lambda) \non  \\
&+& \sum_{m=1}^{\n-1} \sum_{\alpha=1}^{\nu^{(m)}}
\left[ \delta(\lambda - \tilde\lambda_{\alpha}^{(m)}) \delta_{j,m}
- {\cal R}_{jm}(\lambda - \tilde\lambda_{\alpha}^{(m)}) 
+( \tilde\lambda_{\alpha}^{(m)} \rightarrow 
- \tilde\lambda_{\alpha}^{(m)} ) \right] \Big\} \,,
\label{sigmaopen}
\ee 
where the quantity $\delta \sigma^{(j)}_{(l)}(\lambda)$ defined by
\be 
\delta \sigma^{(j)}_{(l)}(\lambda) = {1\over N} 
\left( {\cal R}_{jl} * \left( - a_{2\xi_{-} + l}
+ a_{2\xi_{+} - 2 \n + l} \right) \right) (\lambda)
\label{deltasigma}
\ee
has the dependence on the boundary parameters $\xi_{\mp}$, and 
${\cal R}_{jm}(\lambda)$ is the resolvent, which has the Fourier transform 
(\ref{inverse}).

We shall also need the densities $\sigma'^{(j)}_{(l)\ s}(\lambda)$
corresponding to the dual pseudovacuum $\omega'$ given by Eq. (\ref{dualvac})
in order to calculate boundary $S$ matrices. According to the 
Lemma, the BAE with the pseudovacuum $\omega'$ are given by Eq. 
(\ref{openBAE}), except with $\xi_{\mp} 
\rightarrow \xi'_{\mp}$ and $l \rightarrow l'$. It follows that the 
corresponding densities $\sigma'^{(j)}_{(l)\ s}(\lambda)$ are given 
by Eq. (\ref{sigmaopen}), except with 
\be
\delta \sigma'^{(j)}_{(l)}(\lambda) = {1\over N} 
\left( {\cal R}_{j,\n -l} * \left(  a_{2\xi_{-} - \n + l}
- a_{2\xi_{+} - \n + l} \right) \right) (\lambda)
\label{deltasigmaprime}
\,. 
\ee

As previously noted, the $K$ matrices in the transfer matrix 
$t_{(l)}(\lambda\,, \xi_{\mp})$ break the bulk $SU(\n)$ symmetry.  
Hence, strictly speaking, states should be classified according to the 
unbroken symmetry $SU(l) \times SU(\n-l) \times U(1)$.  However, we 
expect that at points of the chain that are far from the boundary, the 
effects of the boundary should be ``small''.  In particular, in the 
bulk, multiparticle states should ``approximately'' form irreducible 
representations of $SU(\n)$, as discussed for the closed chain in 
Section 2.  Therefore, we shall continue to classify bulk 
multiparticle states by $SU(\n)$ quantum numbers.

Consider now the special case of one-particle states of type $[1]$, 
which form an $\n$-dimensional representation.  For the Bethe Ansatz 
state constructed with the pseudovacuum $\omega_{(k)}$ having one hole 
in sea 1, the Cartan generators have the eigenvalues \footnote{In 
particular, $S^{(j)} = \delta_{j,1}$ for $k=1$, which is consistent 
with Eq.  (\ref{nice}).}
\be
S^{(j)} = \delta_{j,k} - \delta_{j,k-1} \,.
\label{cart1}
\ee 
This state is represented by the vector
\be
\left( \begin{array}{c}
                       0\\
                       \vdots \\
                       1 \\
                       \vdots \\
                       0
                       \end{array} \right)
                       \vphantom{\begin{array}{c}
                       0\\
                       \vdots \\
                       1 \\
                       \end{array}} \leftarrow k^{th} \,, 
\ee 
which is the eigenvector of the matrices $\{ s^{(j)} \}$ given in Eq.  
(\ref{cartan}) with the eigenvalues (\ref{cart1}).   

Similarly, the one-particle states of type $[\n-1] = [\bar 1]$ also 
form an $\n$-dimensional representation.  The Bethe Ansatz state 
constructed with the pseudovacuum $\omega_{(k)}$ having one hole in 
sea $\n -1$ has the eigenvalues
\be
S^{(j)} = \delta_{j,\n - k} - \delta_{j,\n + 1 - k} \,.
\label{cart2}
\ee 
This state is represented by the vector
\be
\left( \begin{array}{c}
                       0\\
                       \vdots \\
                       1 \\
                       \vdots \\
                       0
                       \end{array} \right)
                       \vphantom{\begin{array}{c}
                       0\\
                       \vdots \\
                       1 \\
                       \end{array}} \leftarrow 
                       \left( \n + 1 - k \right)^{th} \,, 
\label{decart}                       
\ee 
keeping in mind that the Cartan generators are now represented by the 
matrices $\{ -s^{(j)*} \}$. The highest weight of $[\bar 1]$ is the 
negative of the lowest weight of $[1]$.

\subsection{Boundary $S$ matrices}

We define the boundary $S$ matrices $S^{\mp}_{(l)\ [j]}$ for a particle 
of type $[j]$, in analogy with the bulk $S$ matrix, by the 
quantization condition \cite{GMN}, \cite{DMN}
\be
\left( e^{i 2 p^{(j)}(\tilde\lambda^{(j)}) N} 
S^{+}_{(l)\ [j]}\ S^{-}_{(l)\ [j]}- 1 \right) 
|\tilde\lambda^{(j)} \rangle = 0 \,,
\label{quantizationopen}
\ee 
where $p^{(j)}(\lambda)$ is defined by Eq. (\ref{needlater}), 
and $\tilde\lambda^{(j)}$ is the hole rapidity.
There is an identity for the open chain which is analogous to the one 
given in Eq.  (\ref{identity}) for the closed chain:
\be 
{1\over \pi} {d\over d\lambda}p^{(j)}(\lambda) 
+ \sigma^{(j)}_{(l)}(\lambda) - 2 s^{(j)}(\lambda)  = 
{1\over N} {d\over d\lambda}h^{(j)}_{(l)}(\lambda) 
\ee
For simplicity, we focus our attention on the cases $[1]$ and $[\n-1] 
= [\bar 1]$, for which the $S$ matrices act in $C_{\n}$ and 
$\overline{C_{\n}}$, respectively.

We first treat the case $[1]$. The $SU(l) \times SU(\n-l) \times 
U(1)$ invariance of the transfer matrix implies that $S^{\mp}_{(l)\ 
[1]}$ are diagonal $\n \times \n$ matrices of the form
\be
S^{\mp}_{(l)\ [1]} = diag \Bigl(  
\underbrace{\alpha^{\mp}_{(l)}\,, \ldots \,, \alpha^{\mp}_{(l)}}_{l} \,, 
\underbrace{\beta^{\mp}_{(l)}\,, \ldots \,, \beta^{\mp}_{(l)}}_{\n-l} \Bigr) \,.
\label{form[1]}
\ee
Choosing the state $|\tilde\lambda^{(1)} \rangle$ in 
Eq. (\ref{quantizationopen}) to be the Bethe Ansatz state 
constructed with the pseudovacuum $\omega_{(1)}$ having one hole in 
sea 1, we see that (up to a rapidity-independent phase factor) 
\be
\alpha^{+}_{(l)}\ \alpha^{-}_{(l)}  \sim   
\exp \left\{ i 2\pi N \int_{0}^{\tilde\lambda^{(1)}}
\left( \sigma_{(l)}^{(1)}(\lambda) - 2 s^{(1)}(\lambda) \right) d\lambda
\right\} \,,
\label{openrel}
\ee 
where $\sigma_{(l)}^{(1)}(\lambda)$ is given by Eq. (\ref{sigmaopen}).
Using the fact
\be
\int_{0}^{\tilde\lambda^{(1)}} \left[ 
{\cal R}\left(\lambda - \tilde\lambda^{(1)}\right) 
+ {\cal R}\left(\lambda + \tilde\lambda^{(1)}\right) \right]\ d\lambda 
= \int_{0}^{\tilde\lambda^{(1)}} 2 {\cal R}\left(2 \lambda \right)\ d\lambda 
\,,
\ee
and with the help of the identity (\ref{GRintegral}) and the duplication formula
for the gamma function
\be
2^{2x-1}\Gamma \left( x \right) \Gamma \left( x + {1\over 2} \right) =
\pi^{1\over 2} \Gamma \left( 2 x \right) \,,
\ee 
we find
\be
\alpha^{-}_{(l)} &=& S_{0}(\tilde\lambda^{(1)}) 
{\Gamma \left({1\over \n}\left(\xi_{-} + l - {1\over 2}
+ i \tilde\lambda^{(1)} \right) \right) 
\Gamma \left({1\over \n}\left(\xi_{-} + \n - {1\over 2}
- i \tilde\lambda^{(1)} \right) \right)\over
\Gamma \left({1\over \n}\left(\xi_{-} + l - {1\over 2}
- i \tilde\lambda^{(1)} \right) \right) 
\Gamma \left({1\over \n}\left(\xi_{-} + \n - {1\over 2}
+ i \tilde\lambda^{(1)} \right) \right)} \,, \non \\
\alpha^{+}_{(l)} &=& S_{0}(\tilde\lambda^{(1)}) 
{\Gamma \left({1\over \n}\left(\xi_{+} - \n + l - {1\over 2}
- i \tilde\lambda^{(1)} \right) \right) 
\Gamma \left({1\over \n}\left(\xi_{+} - {1\over 2}
+ i \tilde\lambda^{(1)} \right) \right)\over
\Gamma \left({1\over \n}\left(\xi_{+} - \n + l - {1\over 2}
+ i \tilde\lambda^{(1)} \right) \right) 
\Gamma \left({1\over \n}\left(\xi_{+} - {1\over 2}
- i \tilde\lambda^{(1)} \right) \right)} \,,
\label{alphas}
\ee
where the prefactor $S_{0}(\tilde\lambda)$ is given by
\be
S_{0}(\tilde\lambda) = {\Gamma \left({1\over \n}\left({1\over 2}(\n -1) 
- i \tilde\lambda \right) \right) 
\Gamma \left({1\over \n}\left(\n + i \tilde\lambda \right) \right)\over
\Gamma \left({1\over \n}\left({1\over 2}(\n -1) 
+ i \tilde\lambda \right) \right) 
\Gamma \left({1\over \n}\left(\n - i \tilde\lambda \right) \right)}
\,. \label{prefactor}
\ee 
Moreover, choosing the state $|\tilde\lambda^{(1)} \rangle$ in Eq.  
(\ref{quantizationopen}) to be the Bethe Ansatz state constructed with 
the dual pseudovacuum $\omega' = \omega_{(\n)}$ having one hole in sea 
1, we obtain the relation
\be
{\beta^{+}_{(l)}\ \beta^{-}_{(l)}\over \alpha^{+}_{(l)}\ 
\alpha^{-}_{(l)}} = 
\exp \left\{ i 2\pi N \int_{0}^{\tilde\lambda^{(1)}}
\left( \sigma'^{(1)}_{(l)}(\lambda) - \sigma^{(1)}_{(l)}(\lambda)
\right) d\lambda \right\}  \,.
\ee 
Note that
\be
\sigma'^{(1)}_{(l)}(\lambda) - \sigma^{(1)}_{(l)}(\lambda) &=&
\delta\sigma'^{(1)}_{(l)}(\lambda) - 
\delta\sigma^{(1)}_{(l)}(\lambda) \non \\ 
&=& {1\over N} \left( a_{2\xi_{-}-1}(\lambda) 
- a_{2\xi_{+} - 2\n + 2l - 1}(\lambda) \right) \,, 
\ee 
where $\delta\sigma^{(1)}_{(l)}$ and $\delta\sigma'^{(1)}_{(l)}$ are 
given by Eqs.  (\ref{deltasigma}) and (\ref{deltasigmaprime}), 
respectively. We conclude
\be
{\alpha^{-}_{(l)}\over \beta^{-}_{(l)}} &=& 
-e_{2\xi_{-} - 1}(\tilde\lambda^{(1)}) \,, \non \\ 
{\beta^{+}_{(l)}\over \alpha^{+}_{(l)}} &=& 
-e_{2\xi_{+} - 2\n + 2l - 1}(\tilde\lambda^{(1)}) \,,
\label{betas}
\ee
where we have resolved the sign ambiguity by demanding that the $S$ 
matrix be proportional to the unit matrix for $\tilde\lambda^{(1)} = 0$.

Finally, we consider the case $[\n-1]$. The boundary $S$ matrices 
$S^{\mp}_{(l)\ [\n-1]}$ are diagonal $\n \times \n$ matrices of the form
\be
S^{\mp}_{(l)\ [\n-1]} = diag \Bigl(  
\underbrace{\bar \alpha^{\mp}_{(l)}\,, \ldots \,, \bar \alpha^{\mp}_{(l)}}_{l} \,, 
\underbrace{\bar \beta^{\mp}_{(l)}\,, \ldots \,, \bar \beta^{\mp}_{(l)}}_{\n-l} 
\Bigr) \,.
\label{form[n-1]}
\ee
For this case we must consider one hole in sea $\n - 1$. Noting that 
\be
\bar\beta^{+}_{(l)}\ \bar\beta^{-}_{(l)} \sim   
\exp \left\{ i 2\pi N \int_{0}^{\tilde\lambda^{(\n -1)}}
\left( \sigma_{(l)}^{(\n -1)}(\lambda) - 2 s^{(\n -1)}(\lambda) \right) d\lambda
\right\} \,,
\ee 
(see Eq. (\ref{decart})), we obtain
\be
\bar\beta^{-}_{(l)} &=& S_{0}(\tilde\lambda^{(\n -1)}) 
{\Gamma \left({1\over \n}\left(\xi_{-} + l + {1\over 2}(\n -1)
- i \tilde\lambda^{(\n -1)} \right) \right)
\Gamma \left({1\over \n}\left(\xi_{-} + {1\over 2}(\n -1)
+ i \tilde\lambda^{(\n -1)} \right) \right) \over
\Gamma \left({1\over \n}\left(\xi_{-} + l + {1\over 2}(\n -1)
+ i \tilde\lambda^{(\n -1)} \right) \right)
\Gamma \left({1\over \n}\left(\xi_{-} + {1\over 2}(\n -1)
- i \tilde\lambda^{(\n -1)} \right) \right)} \,, \non \\
\bar\beta^{+}_{(l)} &=& S_{0}(\tilde\lambda^{(\n -1)}) 
{\Gamma \left({1\over \n}\left(\xi_{+} + l - {1\over 2}(\n +1)
+ i \tilde\lambda^{(\n -1)} \right) \right)
\Gamma \left({1\over \n}\left(\xi_{+} - {1\over 2}(\n +1)
- i \tilde\lambda^{(\n -1)} \right) \right) \over
\Gamma \left({1\over \n}\left(\xi_{+} + l - {1\over 2}(\n +1)
- i \tilde\lambda^{(\n -1)} \right) \right)
\Gamma \left({1\over \n}\left(\xi_{+} - {1\over 2}(\n +1)
+ i \tilde\lambda^{(\n -1)} \right) \right)} \,, \non \\
& & 
\label{baralphas}
\ee
where $S_{0}(\tilde\lambda) $ is given in Eq. (\ref{prefactor}). 
Moreover,
\be
{\bar\beta^{-}_{(l)}\over \bar\alpha^{-}_{(l)}} &=& 
-e_{2\xi_{-} + 2l - \n - 1}(\tilde\lambda^{(\n -1)}) \,, \non \\ 
{\bar\alpha^{+}_{(l)}\over \bar\beta^{+}_{(l)}} &=& 
-e_{2\xi_{+} - \n - 1}(\tilde\lambda^{(\n -1)}) \,.
\ee

\section{Discussion}

We have shown how to describe general multiparticle states of the 
antiferromagnetic $SU(\n)$ chain, in particular their $SU(\n)$ quantum 
numbers, within the framework of the Bethe Ansatz/string hypothesis.  
The picture which emerges is a rich generalization of the $\n=2$ case 
\cite{FT}, \cite{faddeev/takhtajan}.  The ubiquitous appearance of the 
kernel $\left( 1 + \hat {\cal K}(\omega)\right)_{jm}$, which is 
characterized (see, e.g.,  \cite{devega}) by the $SU(\n)$ Dynkin 
diagram, is noteworthy.  Moreover, we have computed both bulk and 
boundary scattering matrices for particles of types $[1]$ and $[\n 
-1]$.  It should be possible to extend this analysis to particles of 
any type $[k]$.

We have identified a ``duality'' symmetry of the open chain transfer 
matrix with diagonal boundary fields, which plays an important role in 
our computation of boundary $S$ matrices.  It may be interesting to 
investigate further the ``self-dual'' case.  We expect that this 
symmetry is also present for the boundary $A_{\n-1}^{(1)}$ Toda 
theories with diagonal boundary fields. Whether such symmetries 
persist for nondiagonal boundary fields is also an interesting 
question. 

The ``mixed'' boundary condition case $l_{+} \ne l_{-}$ merits further 
investigation.  One expects that the boundary $S$ matrix for one end 
of the chain should be independent of the boundary conditions at the 
other end.  However, for this case, the unbroken symmetry group of the 
transfer matrix is smaller, and therefore, the arguments presented here 
require further refinement.

Since the groups $SO(2\n)$ and $E_{\n}$, like $SU(\n)$, have 
simply-laced Lie algebras, it should be possible to treat the 
corresponding integrable chains in a similar fashion.  Finally, we 
note that the boundary $S$ matrix calculations presented here can be 
generalized \cite{DN} to the trigonometric case, i.e., the open chain 
constructed with the $A_{\n-1}^{(1)}$ $R$ and $K$ matrices.

\section{Acknowledgments}

We are grateful to O. Alvarez and L. Mezincescu for valuable 
discussions.  This work was supported in part by the National Science 
Foundation under Grant PHY-9507829.

\vfill\eject

\end{document}